\documentclass[11pt,english]{article}
\usepackage[latin9]{inputenc}
\usepackage{amsmath}
\usepackage{amssymb}
\usepackage{cancel}
\usepackage{graphicx}
\usepackage{setspace}
\makeatletter

\providecommand{\tabularnewline}{\\}

\numberwithin{equation}{section}

\@ifundefined{date}{}{\date{}}
\usepackage{esint}
\usepackage{hyperref}
\usepackage{latexsym}
\usepackage{graphicx}\usepackage{bm}\usepackage{longtable}

\usepackage{xcolor}

\@addtoreset{equation}{section}

\makeatother

\usepackage{babel}
\begin{document}
\title{Worldvolume fermion as baryon with homogeneous instantons in holographic
QCD\textsubscript{3}}
\maketitle
\begin{center}
Si-wen Li\footnote{Email: siwenli@dlmu.edu.cn}, Xiao-tong Zhang, 
\par\end{center}

\begin{center}
\emph{Department of Physics, School of Science,}\\
\emph{ Dalian Maritime University,}\\
\emph{ Dalian 116026, China}
\par\end{center}

\vspace{12mm}

\begin{abstract}
We investigate holographically the effective theory of the worldvolume
fermion on the flavor branes in the D3/D7 model with homogeneously
smeared D(-1)-branes. As a top-down approach in gauge-gravity duality,
the D(-1)-branes are instantons and violate the CP symmetry in the
dual theory. The background geometry of this model contains black
brane (deconfined geometry) and bubble D3-brane solutions (confined
geometry), all with a non-zero Romand-Romand zero form as axion. The
dual theories to the backgrounds are respectively the Super Yang-Mills
theory at finite temperature and three-dimensional confining Yang-Mills
theory, all with a Chern-Simons term induced by instantons. In the
confined geometry, we introduce a baryon vertex as a D5-brane wrapped
on $S^{5}$, then identify the fermionic fluxes produced by the $N_{c}$
open strings on the D7-brane as a baryonic operator. Afterwards, we
study the spectrum and the holographic correlation function of the
baryonic fermion on the D7-branes. Remarkably, the fermionic spectrum
is in agreement with the dispersion curves obtained from the confined
correlation function, and the mass ratio of the lowest baryon and
meson in our model is close to the associated experimental data. Moreover,
the effective interaction terms of the holographic baryon and meson
are derived and all the coupling constants take order of $N_{c}^{1/2}$
agreeing with the evaluation from the large $N_{c}$ field theory.
In the deconfined geometry, the holographic correlation function is
also evaluated numerically while the fermion on D7-brane is identified
to plasmino instead of baryon. The dispersion curves from the deconfined
correlation function basically covers the results from the hard thermal
loop approximation and may imply the instanton-induced interaction
with spin. Overall, this work constructs a holographic theory about
baryonic fermion and mesonic boson with instantons or CP violation.
\end{abstract}
\newpage{}

\tableofcontents{}

\section{Introduction}

There have been a long history for the investigation of the physics
with CP violation. For instance, A.Sakharov in 1967 \cite{key-1}
proposed that C and CP violation is very necessary in the early universe,
because matter and antimatter were created in equal amounts at the
Big Bang \cite{key-2}, and they would be annihilated in pairs without
matter remaining in today's universe if C and CP symmetry are strictly
preserved. The mechanism of CP violation according to the Standard
Model of particle physics arises from the nonzero phase parameter
of the Cabibbo--Kobayashi--Maskawa (CKM) matrix which describes
successfully the various decays of meson and baryon with CP violation
\cite{key-3,key-4}. Nevertheless, matter-antimatter asymmetry predicted
by the CKM mechanism is very smaller than some astronomical observations
\cite{key-5}, so the discovery of physics beyond the Standard Model
ongoing exploration of CP violation is also significant.

To understand the various phenomena with the hadronic matter-antimatter
asymmetry, the CP violation in strong interaction becomes an theory
of candidate since, in non-Abelian gauge theory, the CP violation
can originate from the topologically non-trivial excitations of the
vacuum, i.e. the instanton \cite{key-6,key-7}. Instantons are consist
of constituents called BPS monopoles or dyons and they are related
to efforts in linking confinement contributing to the thermodynamics
and chiral symmetry breaking \cite{key-8,key-9,key-10,key-11,key-12,key-13,key-14,key-15,key-16,key-17}.
On the other hand, meson and baryon are low-energy confined states
in quantum chromodynamics (QCD) which is very challenging to be described
analytically by using the perturbed method of quantum field theory
(QFT), especially in the presence of instantons or CP violation. Therefore,
in this work, we focus on the top-down approach of the D3/D7 model
\cite{key-18,key-19,key-20,key-21,key-22} with instantons through
the gauge-gravity duality, in order to construct a holographic theory
involving baryon and meson with instantons or CP violation. Our motivations
are as follows. First, gauge-gravity duality is possible to analyze
a gauge field theory in strong coupling limit through its dual classical
gravity theory \cite{key-23,key-24,key-25}, in particular, meson
and baryon can be included naturally in the top-down holographic approach
based on the string theory \cite{key-26,key-27,key-28,key-29}. Second,
instantons in gauge theory are holographically dual to the D(-1)-branes
in the D3/D7 model \cite{key-18,key-28}, and the corresponding geometric
background involving the backreaction of the D(-1)-branes is deformed
$\mathrm{Ad}\mathrm{S}_{5}\times S^{5}$ which can be obtained analytically
by turning on homogeneously the Ramond-Ramond (R-R) zero form $C_{0}$
in the supergravity action \cite{key-18,key-19}. Remarkably, by taking
into account a probe D3-brane located at the holographic boundary
of the bulk, there would be a Chern-Simons (CS) term in the D3-brane
action as \cite{key-30},

\begin{equation}
S_{\mathrm{CS}}\sim\int C_{0}f\wedge f\sim i\frac{\theta}{64\pi^{2}}\int f\wedge f\sim\theta\int\omega_{3},
\end{equation}
plays the role violating the CP symmetry in the dual field theory.
Here $f$ refers to the gauge field strength on the probe D3-brane,
$\omega_{3}$ is the CS 3-form and $\theta$ is the parameter describing
the topological charge carried by the instantonic vacuum. Third, when
the baryon vertex as a D5-brane wrapped on $S^{5}$ (with $N_{c}$
open strings end on it) is introduced into the D3/D7 model, the fermionic
flux on the D7-brane can be a bulk baryonic operator because it is
a gauge-invariant operator combined by $N_{c}$ open strings (as $N_{c}$
quarks) taking baryon numbers \cite{key-29}. Since baryon with instantons
leads to several effects which are very difficult to be analyzed analytically
in QFT \cite{key-6,key-7}, gauge-gravity duality may provide an alternative
method to this goal. Altogether, the D3/D7 model with D(-1)-branes
contains all elements about meson, baryon and CP violation in holography,
thus it would be a novel approach to construct a holographic theory
involving meson and baryon with instantons or CP violation by using
this model.

Let us reveal the outline of this work here. In Section 2, we collect
the essential parts of the D3-brane background with D(-1)-branes as
instantons, then discuss the embedding configuration of the D7-branes
as flavors and the holographic duality for baryon. The remarkable
point here is we can obtain a three-dimensional (3d) confining dual
theory\footnote{The D3/D7 model is also used to study the 3d Yang-Mills Chern-Simons
theory \cite{key-21,key-22,key-a1,key-a2}.} as QCD{\scriptsize 3} by following the classical method in \cite{key-27,key-28,key-31},
and the fermionic flux on the worldvolume of the D7-branes becomes
baryonic field when the baryon vertex with $N_{c}$ open strings is
introduced \cite{key-29,key-32,key-33,key-34}. In Section 3, we study
the mass spectrum, the holographic correlation function of the fermions
on the D7-brane. The onset mass given by the correlation function
and given by the decomposition of the worldvolume fermion are consistent
in the confined phase. In addition, we find the mass ratio of the
lowest baryon and meson in our model is close to the experimental
data. In the deconfined phase, the correlation function basically
covers the dispersion curves obtained from the hard thermal loop (HTL)
approximation in thermal field theory\footnote{Searching for the theta vacuum or instantons in hot QCD also attracts
great interests \cite{key-a3,key-a4}.}. In Section 4, we demonstrate briefly how to derive the interaction
terms involving baryonic fermion and mesonic boson from the D7-brane
action. It indicates that the CP symmetry may not be necessary in
the baryon decay. The summary and discussion are given in the last
section. In the appendix, we investigate the spectroscopy of the mesonic
boson on the D7-brane which is useful to evaluate numerically the
coupling terms of the baryonic fermion and mesonic boson in the low-energy
effective action.

\section{Setup of the D3/D7 model with homogenous instanton}

In this section, let us briefly review the D3/D7 model with homogenous
instantons as D(-1)-branes, then focus on the holographic correspondence
with respect to baryon.

\subsection{The D3-brane background with D-instanton}

The background geometry of the D3/D7 model with homogenous instantons
is created by $N_{c}$ D3-branes with $N_{\mathrm{D}}$ D-instantons,
i.e. the $N_{\mathrm{D}}$ D(-1)-branes, in the large-$N_{c}$ limit
\cite{key-18,key-19}. Its supergravity solution is in general a 10d
deformed D3-brane solution with a non-trivial Ramond-Ramond (R-R)
zero form $C_{0}$ which is recognized as a marginal \textquotedblleft bound
state\textquotedblright{} of D3-branes with homogeneously smeared
$N_{\mathrm{D}}$ D(-1)-branes. In our notation, the $N_{c}$ D3-branes
are identified as color branes, hence its number $N_{c}$ denotes
the color number. Then the low-energy dynamics of the background is
described by type IIB supergravity action which is given in string
frame as \cite{key-27},

\begin{equation}
S_{\mathrm{IIB}}=\frac{1}{2\kappa_{10}^{2}}\int d^{10}x\sqrt{-g}\left[e^{-2\Phi}\left(\mathcal{R}+4\partial\Phi\cdot\partial\Phi\right)-\frac{1}{2}\left|F_{1}\right|^{2}-\frac{1}{2}\left|F_{5}\right|^{2}\right],
\end{equation}
where $2\kappa_{10}^{2}=\left(2\pi\right)^{7}l_{s}^{8}$ is the 10d
gravity coupling constant, $l_{s},g_{s}$ is respectively the string
length and the string coupling constant. We use $\Phi$ to denote
the dilaton field and $F_{1,5}$ is the field strength of the R-R
zero and four form $C_{0,4}$ respectively. The solution of $N_{c}$
D3-branes with $N_{\mathrm{D}}$ D-instantons is the near-horizon
solution of non-extremal black D3-branes with a non-trivial $C_{0}$,
which in string frame reads \cite{key-20,key-21,key-35},

\begin{align}
ds^{2} & =e^{\frac{\phi}{2}}\left\{ \frac{r^{2}}{R^{2}}\left[-f\left(r\right)dt^{2}+d\mathbf{x}\cdot d\mathbf{x}\right]+\frac{1}{f\left(r\right)}\frac{R^{2}}{r^{2}}dr^{2}+R^{2}d\Omega_{5}^{2}\right\} ,\nonumber \\
e^{\phi} & =1+\frac{Q}{r_{H}^{4}}\ln\frac{1}{f\left(r\right)},\ f\left(r\right)=1-\frac{r_{H}^{4}}{r^{4}},\ F_{5}=dC_{4}=g_{s}^{-1}\mathcal{Q}_{3}\epsilon_{5},\nonumber \\
F_{1} & =dC_{0},\ C_{0}=-ie^{-\phi}+i\mathcal{C},\ \phi=\Phi-\Phi_{0},\ e^{\Phi_{0}}=g_{s}.\label{eq:2.2}
\end{align}
where $\mathcal{C}$ is a boundary constant for $C_{0}$, $\epsilon_{5}$
is the volume element of a unit $S^{5}$. And the associated parameters
are given as,

\begin{equation}
R^{4}=4\pi g_{s}N_{c}l_{s}^{4},\ \mathcal{Q}_{3}=4R^{4}g_{s},\ Q=\frac{N_{\mathrm{D}}}{N_{c}}\frac{\left(2\pi\right)^{4}\alpha^{\prime2}}{V_{4}}\mathcal{Q}_{3}.
\end{equation}
Here, $x^{\mu}=\left\{ t,\mathbf{x}\right\} =\left\{ t,x^{i}\right\} ,\ i=1,2,3$
refers to the 4d spacetime where the D3-branes are extended along.
And $r$ is the holographic direction perpendicular to the D3-branes.
The solution (\ref{eq:2.2}) is asymptotic $\mathrm{AdS}_{5}\times S^{5}$
at the holographic boundary $r\rightarrow\infty$ which describes
geometrically that the D-instanton charge $N_{\mathrm{D}}$ is smeared
over the worldvolume $V_{4}$ of the $N_{c}$ coincident black D3-branes
homogeneously with a horizon at $r=r_{H}$. The backreaction of the
D-instantons has been taken into account in the background, thus it
implies $N_{\mathrm{D}}/N_{c}$ must be fixed in the large-$N_{c}$
limit. The dual theory of this background is conjectured as the four
dimensional (4d) $\mathcal{N}=4$ super Yang-Mills theory (SYM) in
a self-dual gauge field (instantonic) background or with a dynamical
axion at finite temperature characterized by the parameter $Q$ \cite{key-20,key-21,key-35}.
Note that, in the language of hadron physics, $C_{0}$ is recognized
as axion and the gluon condensate in this system can be evaluated
as,

\begin{equation}
\left\langle \mathrm{Tr}F_{\mu\nu}F^{\mu\nu}\right\rangle \simeq\frac{N_{\mathrm{D}}}{16\pi^{2}V_{4}}=\frac{Q}{\left(2\pi\alpha^{\prime}\right)^{2}R^{4}}\frac{N_{c}}{\left(2\pi\right)^{4}},\ \mu,\nu=0,1...3.\label{eq:2.4}
\end{equation}

In gauge-gravity duality, the black brane background (\ref{eq:2.2})
is holographically dual to the deconfined SYM theory at high temperature
\cite{key-24,key-27}. Nevertheless, it is possible to obtain a confined
geometry from (\ref{eq:2.2}) by following the classical project proposed
in \cite{key-27,key-28,key-31}. Specifically, compactify one of the
three spatial dimensions of the D3-branes on a circle $S^{1}$. Hence,
below the Kaluza-Klein energy scale $M_{KK}=2\pi/\delta x^{i}$, the
dual theory becomes effectively three-dimensional, where $\delta x^{i}$
refers to the period of $S^{1}$. Then the supersymmetric fermions
and scalars in the dual theory will acquire mass of order $M_{KK}$
when the anti-periodic and periodic boundary condition on fermion
and bosonic fields are respectively imposed along $S^{1}$. Therefore,
they are accordingly decoupled in the low-energy dynamics so that
the low-energy effective theory on the compactified D3-branes is purely
3d Yang-Mills theory. Afterwards, perform a double Wick rotation on
the D(-1)-D3 brane background as $t\rightarrow-ix^{i},x^{i}\rightarrow-it$
to identify the bulk gravity solution as its holographic correspondence.
We denote the direction along $S^{1}$ as $x^{3}$ without loss of
generality, thus the confining solution obtained from (\ref{eq:2.2})
is,

\begin{align}
ds^{2} & =e^{\frac{\phi}{2}}\left\{ \frac{r^{2}}{R^{2}}\left[\eta_{\mu\nu}dx^{\mu}dx^{\nu}+f\left(r\right)\left(dx^{3}\right)^{2}\right]+\frac{1}{f\left(r\right)}\frac{R^{2}}{r^{2}}dr^{2}+R^{2}d\Omega_{5}^{2}\right\} ,\nonumber \\
e^{\phi} & =1+\frac{Q}{r_{KK}^{4}}\ln\frac{1}{f\left(r\right)},\ f\left(r\right)=1-\frac{r_{KK}^{4}}{r^{4}},\ F_{5}=dC_{4}=g_{s}^{-1}\mathcal{Q}_{3}\epsilon_{5},\nonumber \\
F_{1} & =dC_{0},\ C_{0}=-ie^{-\phi}+i\mathcal{C},\ \phi=\Phi-\Phi_{0},\ e^{\Phi_{0}}=g_{s}.\label{eq:2.5}
\end{align}
The solution (\ref{eq:2.5}) defined for $r>r_{KK}$ does not have
a horizon since the warp factor $e^{\phi/2}\frac{r^{2}}{R^{2}}$ never
goes to zero. And the asymptotics of the Wilson loop in this geometry
lead to an area law in the dual theory. So it means the dual field
theory exhibit confinement below the energy scale $M_{KK}$ at very
low temperature. In addition, to avoid the conical singularities in
the region of $r>r_{KK}$, we have to further require,

\begin{equation}
M_{KK}=\frac{2r_{KK}}{R^{2}}.
\end{equation}

\subsection{Embedding of the D7-brane}

To construct a dual theory like QCD, the fundamental quark must be
introduced into the dual theory which can be achieved by embedding
a stack of probe $N_{f}$ D7-branes as flavors. In our setup, we require
that the D7-branes are perpendicular to the compactified direction
$x^{3}$, spanning the $\mathbb{R}^{1,2}$ denoted by $\left\{ t,x,y\right\} $,
the holographic direction denoted by $r$ and four of the five directions
in $\Omega_{5}$ as \cite{key-20,key-22}. Accordingly, the fermion
on the D7-branes survives under the dimensional reduction by following
\cite{key-27,key-28,key-31} and, as we will see, it can be identified
to the dual operator of the baryonic field with baryon vertex in the
dual theory. 

To obtain the induced metric on the D7-branes, we impose the following
coordinate transformation

\begin{equation}
r=\frac{r_{H}}{\sqrt{2}}\sqrt{\frac{r_{H}^{2}}{\rho^{2}}+\frac{\rho^{2}}{r_{H}^{2}}},
\end{equation}
so the metric presented in (\ref{eq:2.2}) becomes

\begin{equation}
ds^{2}=e^{\frac{\phi}{2}}\left\{ \frac{r^{2}}{R^{2}}\left[-f\left(r\right)dt^{2}+d\mathbf{x}\cdot d\mathbf{x}\right]+\frac{R^{2}}{\rho^{2}}\left(d\rho^{2}+\rho^{2}d\Omega_{5}^{2}\right)\right\} .\label{eq:2.8}
\end{equation}
We further introduce the coordinate transformation $\zeta=\rho\cos\Theta,w=\rho\sin\Theta$
i.e. $\rho^{2}=w^{2}+\zeta^{2}$ where $\Theta$ is one of the angular
coordinates in $\Omega_{5}$, then the metric (\ref{eq:2.8}) takes
the final form as,

\begin{align}
ds^{2} & =e^{\frac{\phi}{2}}\left\{ \frac{r^{2}}{R^{2}}\left[-f\left(r\right)dt^{2}+d\mathbf{x}\cdot d\mathbf{x}\right]+\frac{R^{2}}{\rho^{2}}\left(d\zeta^{2}+\zeta^{2}d\Omega_{4}^{2}+dw^{2}\right)\right\} .
\end{align}
So the induced metric on the D7-brane can be found by choosing the
simple embedding solution as $x^{3},w=\mathrm{const}$, it leads,

\begin{equation}
ds_{\mathrm{D7}}^{2}=e^{\frac{\phi}{2}}\left\{ \frac{r^{2}}{R^{2}}\left[-f\left(r\right)dt^{2}+dx^{2}+dy^{2}\right]+\frac{R^{2}}{\rho^{2}}\left(d\zeta^{2}+\zeta^{2}d\Omega_{4}^{2}\right)\right\} .\label{eq:2.10}
\end{equation}
Follow the same steps, the confining induced metric can be obtained
from (\ref{eq:2.5}) as,

\begin{equation}
ds_{\mathrm{D7}}^{2}=e^{\frac{\phi}{2}}\left\{ \frac{r^{2}}{R^{2}}\eta_{\mu\nu}dx^{\mu}dx^{\nu}+\frac{R^{2}}{\rho^{2}}\left(d\zeta^{2}+\zeta^{2}d\Omega_{4}^{2}\right)\right\} ,\mu,\nu=0,1,2.\label{eq:2.11}
\end{equation}
So the dual theory in confined geometry becomes 3d QCD (QCD{\scriptsize 3}).
The D-brane configuration in our model is given in Table \ref{tab:1}.
We will use the induced metric (\ref{eq:2.10}) and (\ref{eq:2.11})
in the following discussion for the dual theory in the deconfined
and in the confined phase respectively. To further simplify the calculation,
we will set $w=0$ in this work\footnote{One may worry that $w=0$ causes an issue that the worldvolume of
the D7-branes stops at $r=r_{KK}$ which is forbidden. However this
can be figured out by introducing another (anti) D7-brane stretched
to $r=r_{KK}$ connecting to D7-brane. The same issue can be found
in the $\mathrm{D4/D6/\overline{D6}}$ approach \cite{key-a6} in
which the anti D6-brane is introduced so that the embedding function
can be chosen as zero.}. Since in the D3/D7 model, the boundary value of $w$ is proportional
to the quark mass, it means in our setup, we focus on the massless,
or namely the light flavored, quarks in the dual theory. 
\begin{table}
\begin{centering}
\begin{tabular}{|c|c|c|c|c|c|c|c|}
\hline 
 & $t$ & $x$ & $y$ & $\left(x^{3}\right)$ & $\zeta$ & $\Omega_{4}$ & $w$\tabularnewline
\hline 
\hline 
$N_{\mathrm{D}}$ D(-1)-branes &  &  &  &  &  &  & \tabularnewline
\hline 
$N_{c}$ D3-branes & - & - & - & - &  &  & \tabularnewline
\hline 
$N_{f}$ D7-branes & - & - & - &  & - & - & \tabularnewline
\hline 
Baryon vertex D5-brane &  &  &  &  &  & - & -\tabularnewline
\hline 
\end{tabular}
\par\end{centering}
\caption{\label{tab:1}The D-brane configuration in the D3/D7 model with D(-1)
branes as homogenous instantons. ``-'' represents that the D-brane
extends along this direction. $x^{3}$ is the compactified direction}

\end{table}

\subsection{The holographic correspondence for baryon}

\begin{figure}
\begin{centering}
\includegraphics[scale=0.22]{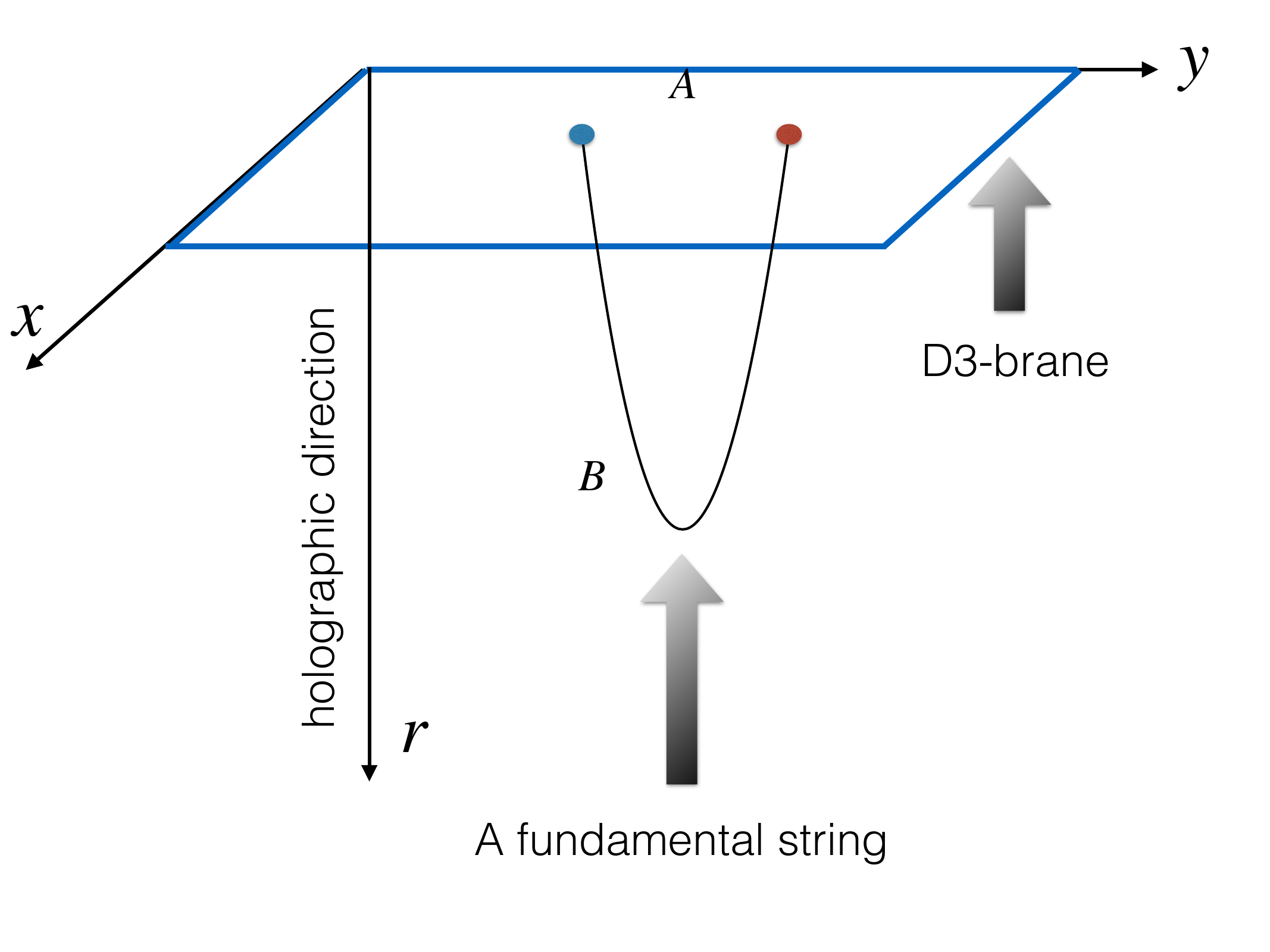}\includegraphics[scale=0.22]{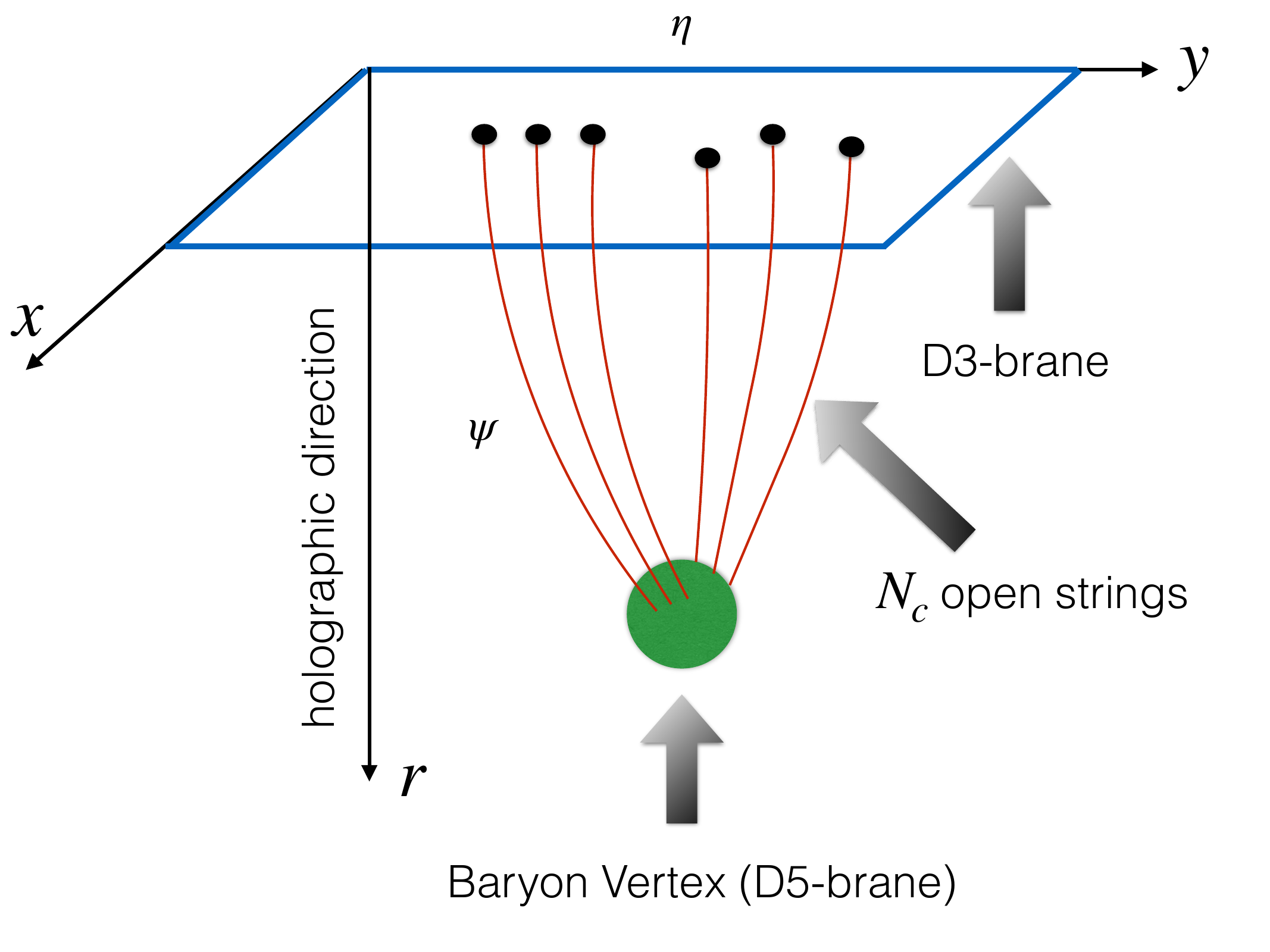}
\par\end{centering}
\caption{\label{fig:1}The D-brane configuration of the D3/D7 system for the
holographic correspondence with respect to meson and baryon. The D7-branes
extend along $x,y,r$ directions. \textbf{The mesonic setup (Left):}
The endpoint (blue) and the anti-endpoint (red) of a fundamental string
refers respectively to a quark and an anti-quark. $A$ is the bosonic
color singlet as a meson consisted of a quark and an anti-quark in
the boundary theory. $B$ is the bosonic flux produced by the fundamental
string as the bulk dual operator to $A$.\textbf{ The baryonic setup
(Right):} The endpoints (black) of the $N_{c}$ strings all with same
orientation are $N_{c}$ quarks at boundary. $\eta$ is the fermionic
color singlet as a baryon consisted of $N_{c}$ quarks at boundary.
$\psi$ is the fermionic flux created by the $N_{c}$ open strings
as the bulk dual operator to $\eta$.}
\end{figure}

In this subsection, let us outline the holographic correspondence
for baryon by taking into account a baryon vertex in our D3/D7 approach.
In gauge-gravity duality, baryon vertex is identified as a probe D-brane
wrapped on the dimensions denoted by the spherical coordinates of
the background geometry with $N_{c}$ open strings ending on it \cite{key-8,key-29,key-a2}.
Accordingly, in the type IIB supergravity on $\mathrm{AdS}_{5}\times S^{5}$,
the baryon vertex is a D5-brane wrapped on $S^{5}$, which is illustrated
in Table \ref{tab:1}, with $N_{c}$ open strings ending on it stretched
to the holographic boundary. Consider a probe D3-brane located at
the holographic boundary, there are $N_{c}$ open strings with same
orientation stretched to the holographic boundary and ended on the
D5-brane as it is illustrated in Figure \ref{fig:1}. The wrapped
D5-brane carries a $U\left(1\right)$ charge proportional to $N_{c}$
due to the contribution from the R-R flux $N_{c}\sim\int_{S^{5}}F_{5}$.
On the other hand, each one of the $N_{c}$ open strings carries a
unit $U\left(1\right)$ Chan-Paton charge $1$ (or $-1$ dependent
on the orientation) which in total cancels exactly the $U\left(1\right)$
charge of the wrapped D5-brane. In this sense, the baryon number,
defined as the $U\left(1\right)$ charge over $N_{c}$, is conserved
on the baryon vertex by choosing suitably the orientation of the D5-brane
and $N_{c}$ open strings \cite{key-8,key-29}.

Then let us specify the holographic duality with respect to baryon.
In the viewpoint of the boundary theory, the endpoints of the $N_{c}$
open strings are inside the D7-branes taking Chan-Paton charge, thus
they are in the fundamental representation of color and flavor group,
namely they are flavored fundamental quarks taking baryon number $1/N_{c}$.
Moreover, there must be possible to form an antisymmetric gauge-invariant
operator $\eta$ as the combination of $N_{c}$ quarks in the boundary
theory, and by taking into account the color confinement, this operator
$\eta$ is the color singlet as a baryon since its baryon number is
one (or minus one for an anti-baryon). Note that this implies the
$N_{c}$ open strings behave as fermion since the color singlet $\eta$
produced by the endpoints of the $N_{c}$ open strings is the antisymmetric
representation of the color group \cite{key-29}. In the bulk theory,
there must be a dual operator to $\eta$ produced by the combination
of the $N_{c}$ open strings according to the holographic principle.
So we denote the $N_{c}$ fluxes in bulk produced by the $N_{c}$
open strings as $\psi$, and $\psi$ also takes the baryon number
as $\eta$. Notice that the constructions of the holographic dualities
for baryon and meson are different. In the mesonic case, the baryon
vertex is not essential since meson is the color singlet consisted
of a quark and an anti-quark, which in holography corresponds to a
fundamental string in the D7-brane ending on the boundary with respect
to both the endpoints, as it is illustrated in Figure \ref{fig:1}.
Therefore, baryon number for a meson is always vanished. 

Afterwards, we need to confirm that QCD baryon in the large $N_{c}$
limit remains to be fermion, in order to find an effective action
for the bulk baryonic field $\psi$. Since baryon is consisted of
$N_{c}$ quarks which are the fundamental representation of the spin
group $SU\left(2\right)$, usually the spin of a baryon can be obtained
by decomposing the tensor product of the $N_{c}$ irreducible representations
of the $SU\left(2\right)$ group for finite $N_{c}$. As a result,
we find baryon is fermion when $N_{c}$ is odd, and it is boson when
$N_{c}$ is even. However, the decomposition of the tensor product
of the $SU\left(2\right)$ group does not have a definite limit at
$N_{c}\rightarrow\infty$, so it is impossible to obtain exactly the
spin of baryon for the case $N_{c}\rightarrow\infty$ in this way.
Fortunately, the spin and flavor group of baryon combine to form the
contracted $SU\left(2N_{f}\right)$ group strictly in the limit of
$N_{c}\rightarrow\infty$ which is the symmetry group of the contracted
spin-flavor symmetry \cite{key-40,key-b2}. Accordingly, baryon in
large $N_{c}$ limit transforms as the irreducible representation
of the group of the spin-flavor symmetry, and the simplest irreducible
representations of the spin-flavor group for the case\footnote{We do not need to consider the case of $N_{f}>3$ in general, since
the flavor number of a baryon usually satisfies $N_{f}\leq3$ in the
realistic QCD.} of $N_{f}=2,3$ are worked out as a tower of states with half-integer
spin and half-integer isospin which illustrates all the baryons are
fermions \cite{key-40,key-b1,key-b2,key-b3}. Therefore, in our holographic
correspondence in the D3/D7 approach, both the baryonic operator $\eta$
in the boundary theory and the dual field $\psi$ in bulk should be
fermionic to match these analysis of baryon in the large $N_{c}$
QCD as the boundary theory. Besides, since the $N_{c}$ open strings
are on the D7-branes, their fermionic fluxes $\psi$ as the bulk baryonic
field can be described effectively by using the fermionic action for
the D7-brane. We note that the analysis in this subsection is valid
only in the confined geometry since there may not be a wrapped configuration
of a D5-brane on $S^{5}$ in the deconfined geometry \cite{key-a5}.
In the rest of this paper, all our discussion in the confined geometry
is in the presence of a baryon vertex, so the $N_{c}$ fermionic fluxes
on the D7-brane in the confined geometry are always baryonic. 

\section{Spectroscopy of the worldvolume fermion on the flavor brane}

As we have specify that the bulk field for a baryon is the fermionic
flux produced by the $N_{c}$ open strings on the D7-branes, in this
section, we study the mass spectrum of the fermion on the D7-brane
in the confined geometry and the correlation function of the fermion
by using the holographic prescription in AdS/CFT with respect to both
the confined and deconfined cases. 

\subsection{The fermionic action of the D-brane}

Since our concern would be the fermionic flux on the flavor brane,
we start from the fermionic action on a $\mathrm{D}_{p}$-brane which
can be obtained by the standard method of T-duality in string theory.
In the type IIB string theory, the quadratic action of worldvolume
fermionic flux $\Psi$ is given as \cite{key-36,key-37},

\begin{equation}
S_{\mathrm{D}_{p}}^{f}=\frac{iT_{p}}{2}\int d^{p+1}xe^{-\phi}\sqrt{-\left(g+\mathcal{F}\right)}\bar{\Psi}\left(1-\Gamma_{\mathrm{D}_{p}}\right)\left(\Gamma^{\alpha}\hat{D}_{\alpha}-\Delta+\mathrm{L}_{\mathrm{D}_{p}}\right)\Psi,\label{eq:3.1}
\end{equation}
where,
\begin{align}
\hat{D}_{\alpha}= & \nabla_{\alpha}+\frac{1}{4\cdot2!}H_{\alpha NK}\Gamma^{NK}\bar{\gamma}-\frac{1}{8}e^{\phi}\big(F_{N}\Gamma^{N}\Gamma_{\alpha}\bar{\gamma}+\frac{1}{3!}F_{KLN}\Gamma^{KLN}\Gamma_{\alpha}\nonumber \\
 & +\frac{1}{2\cdot5!}F_{KLMNP}\Gamma^{KLMNP}\Gamma_{\alpha}\bar{\gamma}\big),\nonumber \\
\Delta= & \frac{1}{2}\left(\Gamma^{M}\partial_{M}\phi+\frac{1}{2\cdot3!}H_{MNK}\Gamma^{MNK}\bar{\gamma}\right)+\frac{1}{2}e^{\phi}\left(F_{M}\Gamma^{M}\bar{\gamma}+\frac{1}{2\cdot3!}F_{KLN}\Gamma^{KLN}\right),\nonumber \\
\Gamma_{\mathrm{D}_{p}}= & \frac{1}{\sqrt{-\left(g+\mathcal{F}\right)}}\sum_{q}\frac{\epsilon^{\alpha_{1}...\alpha_{2q}\beta_{1}...\beta_{p-2q+1}}}{q!2^{q}\left(p-2q+1\right)!}\mathcal{F}_{\alpha_{1}\alpha_{2}}...\mathcal{F}_{\alpha_{2q-1}\alpha_{2q}}\Gamma_{\beta_{1}...\beta_{p-2q+1}}\bar{\gamma}^{\frac{p-2q+1}{2}},\nonumber \\
\mathrm{L}_{\mathrm{D}_{p}} & =\sum_{q}\frac{\epsilon^{\alpha_{1}...\alpha_{2q}\beta_{1}...\beta_{p-2q+1}}}{q!2^{q}\left(p-2q+1\right)!}\frac{\left(-i\sigma_{2}\right)\left(\bar{\gamma}\right)^{\frac{p-2q+1}{2}}}{\sqrt{-\left(g+\mathcal{F}\right)}}\mathcal{F}_{\alpha_{1}\alpha_{2}}...\mathcal{F}_{\alpha_{2q-1}\alpha_{2q}}\Gamma_{\beta_{1}...\beta_{p-2q+1}}^{\ \ \ \ \ \ \ \ \ \ \ \ \ \lambda}\hat{D}_{\lambda}.\label{eq:3.2}
\end{align}
Let us clarify the notation presented in (\ref{eq:3.1}) and (\ref{eq:3.2}).
The indices denoted by capital letters $K,L,M,N...$ run over the
10d spacetime and the indices denoted by lowercase letters $a,b,...$
run over the tangent space of the 10d spacetime. $T_{p}$ refers to
the tension of a $\mathrm{D}_{p}$-brane given as $T_{p}=g_{s}^{-1}\left(2\pi\right)^{-p}l_{s}^{-\left(p+1\right)}$.
The Greek alphabet $\alpha,\beta,\lambda$ refers to the indices running
over the worldvolume of the $\mathrm{D}_{p}$-brane. The metric is
written in terms of vielbein as $g_{MN}=e_{M}^{a}\eta_{ab}e_{N}^{b}$,
and the gamma matrices are given by the relations

\begin{equation}
\left\{ \gamma^{a},\gamma^{b}\right\} =2\eta^{ab},\left\{ \Gamma^{M},\Gamma^{N}\right\} =2g^{MN},
\end{equation}
with $e_{M}^{a}\Gamma^{M}=\gamma^{a}$. Besides, $\omega_{\alpha ab}$
refers to the spin connection and $\nabla_{\alpha}=\partial_{\alpha}+\frac{1}{4}\omega_{\alpha ab}\gamma^{ab}$
is the covariant derivative for fermion. The gamma matrix with multiple
indices is defined by ranking alternate anti-symmetrically or symmetrically
the indices e.g. 
\begin{equation}
\gamma^{ab}=\frac{1}{2}\left[\gamma^{a},\gamma^{b}\right],\gamma^{abc}=\frac{1}{2}\left\{ \gamma^{a},\gamma^{bc}\right\} ,\gamma^{abcd}=\frac{1}{2}\left[\gamma^{a},\gamma^{bcd}\right]...
\end{equation}
$\Gamma^{MNK...}$ shares the same definition as $\gamma^{abc...}$.
$\bar{\gamma}$ is defined as the chiral gamma matrix as $\bar{\gamma}=\gamma^{01...9}$
and $\sigma_{2}$ refers to the associated Pauli matrix in 10d. The
worldvolume field $\mathcal{F}$ is given as $\mathcal{F}=B+\left(2\pi\alpha^{\prime}\right)f$
where $B$ is the NS-NS 2-form in type II string theory with its field
strength $H=dB$ and $f$ is the Yang-Mills field strength. $F_{M},F_{MN},F_{KLM}...$
refer to the field strength of the massless R-R fields presented in
type II string theory. For the D7-brane, let us set $p=7$ then insert
the non-vanished Romand-Romand field and dilaton given in the supergravity
solution (\ref{eq:2.2}) or (\ref{eq:2.5}) into the action (\ref{eq:3.1}),
so the action (\ref{eq:3.1}) becomes,

\begin{align}
S_{\mathrm{D7}}^{f}= & \frac{iT_{7}}{2}\int d^{8}xe^{-\phi}\sqrt{-g}\bar{\Psi}\bigg(\Gamma^{\alpha}\nabla_{\alpha}-\frac{1}{8}e^{\phi}F_{N}\Gamma^{\alpha}\Gamma^{N}\Gamma_{\alpha}\bar{\gamma}\nonumber \\
 & -\frac{1}{2\cdot8\cdot5!}\Gamma^{\alpha}F_{KLMNP}\Gamma^{KLMNP}\Gamma_{\alpha}\bar{\gamma}-\frac{1}{2}\Gamma^{M}\partial_{M}\phi-\frac{1}{2}e^{\phi}F_{M}\Gamma^{M}\bar{\gamma}\bigg)\Psi.\label{eq:3.5}
\end{align}

\subsection{The fermionic spectrum}

\begin{figure}
\begin{centering}
\includegraphics[scale=0.27]{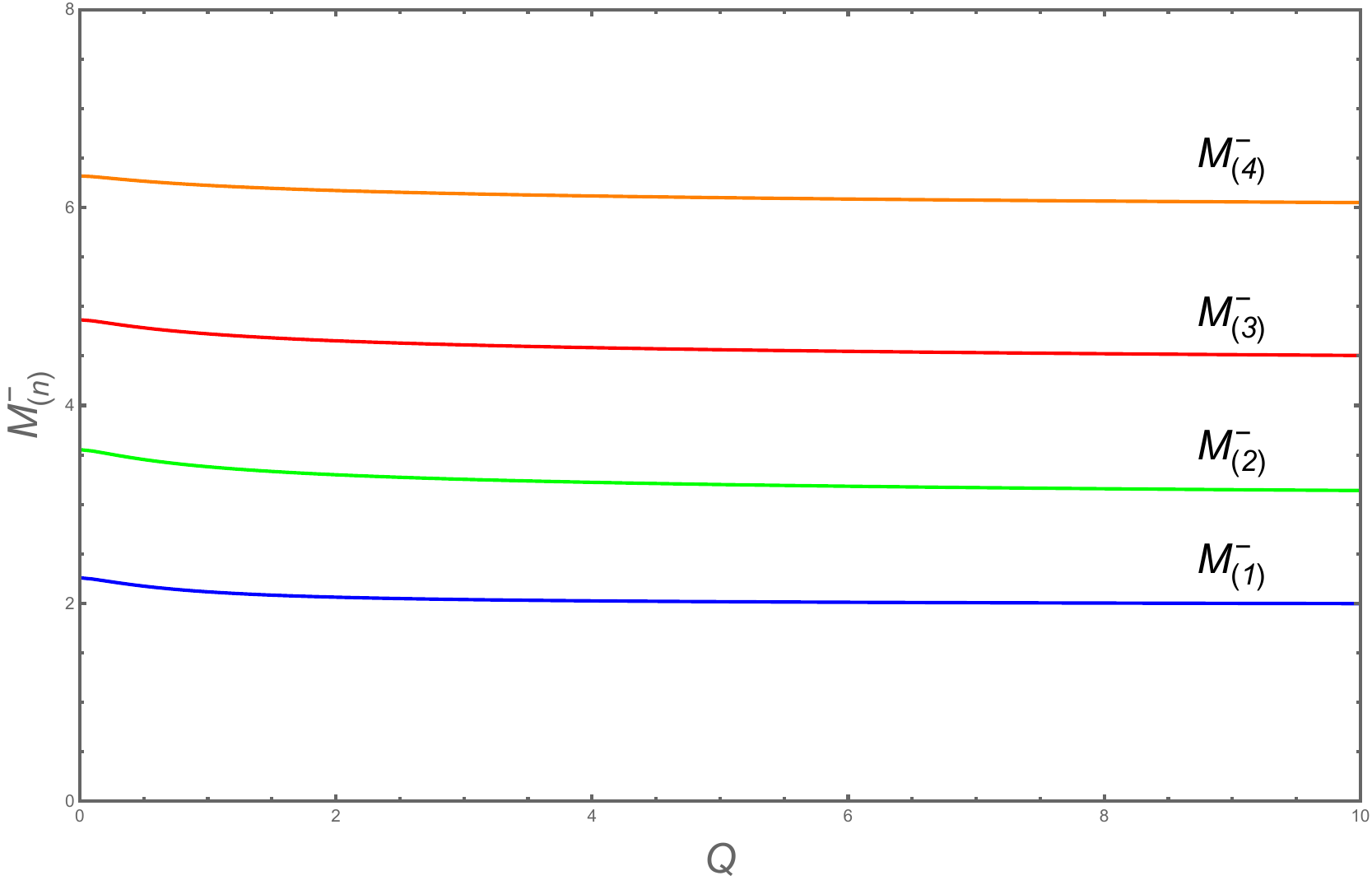}\includegraphics[scale=0.27]{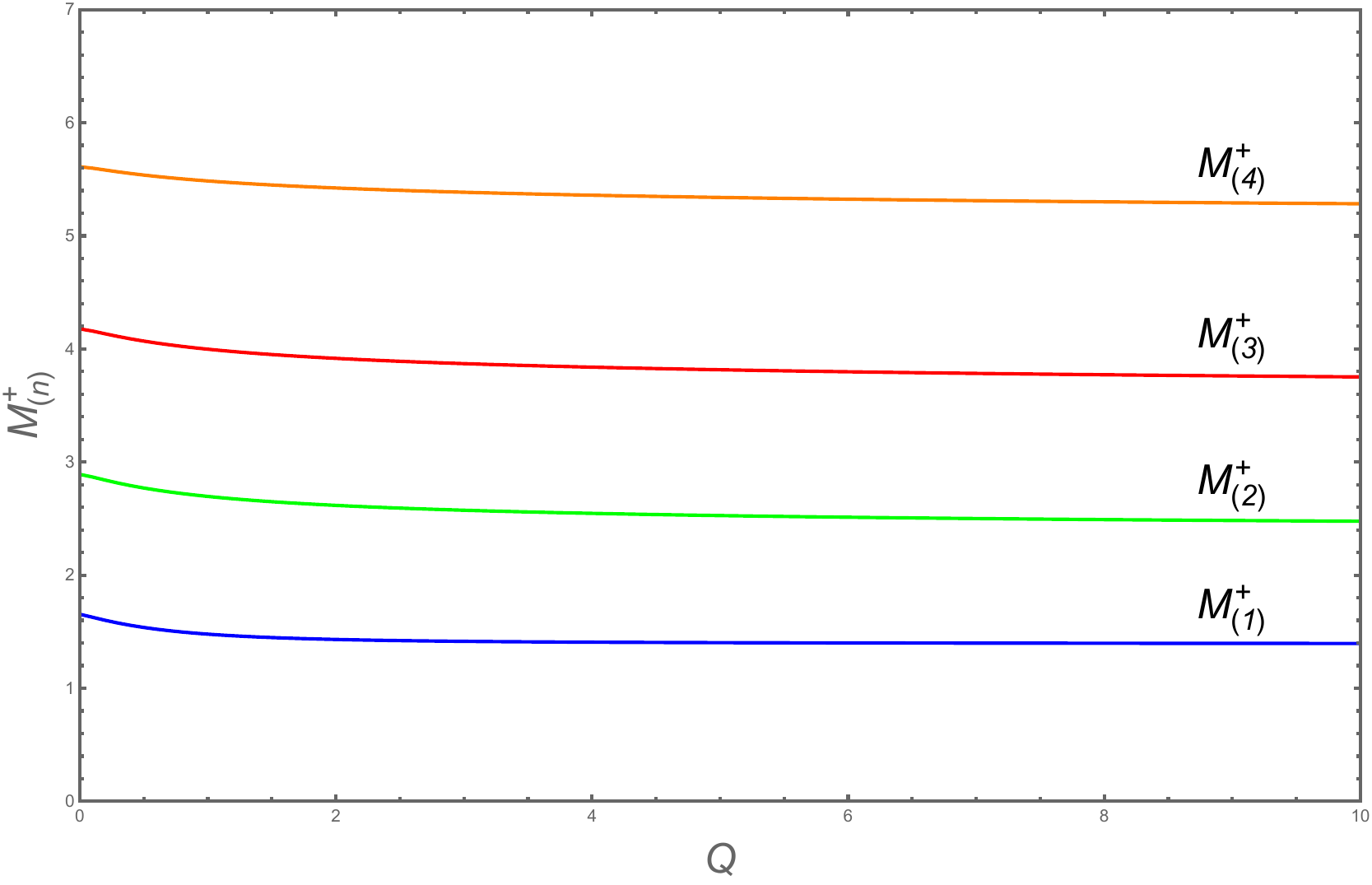}
\par\end{centering}
\caption{\label{fig:2}The dependence on the instanton density $Q$ of the
mass spectrum with $\Lambda_{l}=2$.}

\end{figure}

The holographic fermionic spectrum in the dual theory can be obtained
by integrating out the extra dimensions in the action (\ref{eq:3.5})
as most top-down holographic approaches \cite{key-32,key-33,key-34,key-38}.
Note that the spectrum is only valid in the confined geometry since
its dual theory allows hadronic states exhibiting confinement. To
this goal, let us first calculate the Dirac operator in the confined
geometry (\ref{eq:2.11}) as,

\begin{equation}
\Gamma^{\alpha}\nabla_{\alpha}=e^{-\phi/4}\left(\frac{R}{r}\gamma^{\mu}\partial_{\mu}+\frac{\rho}{R}\gamma^{4}\partial_{4}\right)+\frac{e^{-\phi/4}}{R}\gamma^{4}\left(-\frac{2\zeta}{\rho}+\frac{2\rho}{\zeta}+\frac{3\zeta r^{\prime}}{2r}+\frac{7}{8}\zeta\phi^{\prime}\right)+e^{-\phi/4}\frac{\rho}{R\zeta}\cancel{D}_{S^{4}},\label{eq:3.6}
\end{equation}
where $\cancel{D}_{S^{4}}$ is the Dirac operator on a unit $S^{4}$.
The coordinate on the D7-brane is denoted as $x^{\alpha}=\left\{ t,x,y,x^{4}=\zeta,\Omega_{4}\right\} $
and $x^{\mu}=\left\{ t,x,y\right\} $. In addition, we can further
calculate,

\begin{align}
\frac{1}{8}e^{\phi}F_{N}\Gamma^{\alpha}\Gamma^{N}\Gamma_{\alpha} & =-\frac{3}{4}\frac{\rho}{R}e^{3\phi/4}\partial_{\zeta}C_{0}\gamma^{4},\nonumber \\
\frac{1}{2\cdot8\cdot5!}\Gamma^{\alpha}F_{KLMNP}\Gamma^{KLMNP}\Gamma_{\alpha} & =\frac{1}{2}\frac{e^{-5\phi/4}}{R}\gamma^{56789},\nonumber \\
\frac{1}{2}\Gamma^{M}\partial_{M}\phi+\frac{1}{2}e^{\phi}F_{M}\Gamma^{M}\bar{\gamma} & =\frac{1}{2}\frac{\rho}{R}\gamma^{4}e^{-\phi/4}\left(\partial_{\zeta}\phi+e^{\phi}\partial_{\zeta}C_{0}\bar{\gamma}\right)\label{eq:3.7}
\end{align}
As the kappa symmetry fixes the chirality condition $\bar{\gamma}\Psi=-i\Psi$
in our notation, the 10d spinor can be decomposed into a 3+1 dimensional
part $\psi\left(x,\zeta\right)$ with holographic coordinate $\zeta$,
an $S^{4}$ part $\varphi$ and a remaining 2d part $\beta$ as,

\begin{equation}
\Psi\sim\psi\left(x,\zeta\right)\otimes\varphi\left(S^{4}\right)\otimes\beta,
\end{equation}
satisfying $\gamma^{56789}\Psi=-i\Psi$ and $\beta^{\dagger}\beta=1$.
The spinor $\varphi$ is the eigen function on $S^{4}$ which satisfies
the Dirac equation on $S^{4}$ as \cite{key-39},

\begin{equation}
\cancel{D}_{S^{4}}\varphi=\Lambda_{l}^{\pm}\varphi^{\pm l,s},\ \Lambda_{l}^{\pm}=\pm\left(2+l\right),l=0,1...,\label{eq:3.8}
\end{equation}
and normalization

\begin{equation}
\int\sqrt{g_{S^{4}}}d\Omega_{4}\varphi^{\pm l^{\prime},s^{\prime}\dagger}\varphi^{\pm l,s}=\delta^{l,l^{\prime}}\delta^{s,s^{\prime}},\label{eq:3.10}
\end{equation}
where $s,l$ represent the angular quantum numbers of the spherical
harmonic function. $g_{S^{4}}$ refers to the determinant of the metric
on a unit $S^{4}$. Afterwards, let us insert (\ref{eq:3.6}) - (\ref{eq:3.10})
into action (\ref{eq:3.5}), the fermionic action takes the form as,

\begin{align}
S_{\mathrm{D7}} & =i\int d^{3}xd\zeta d\Omega_{4}\left(\bar{\Psi}\mathcal{A}\gamma^{\mu}\partial_{\mu}\Psi+\bar{\Psi}\mathcal{B}\gamma^{4}\partial_{4}\Psi+\bar{\Psi}\mathcal{C}\gamma^{4}\Psi+\bar{\Psi}\mathcal{D}\Psi\right),\nonumber \\
 & =i\int d^{3}xd\zeta\left[\bar{\psi}\gamma^{\mu}\partial_{\mu}\psi+\bar{\psi}\frac{\mathcal{B}}{\mathcal{A}}\gamma^{4}\partial_{4}\psi+\bar{\psi}\left(\frac{\mathcal{C}}{\mathcal{A}}-\frac{\mathcal{A}^{\prime}\mathcal{B}}{2\mathcal{A}^{2}}\right)\gamma^{4}\psi+\bar{\psi}\frac{\mathcal{D}}{\mathcal{A}}\psi\right]\label{eq:3.11}
\end{align}
where the gamma matrices have been reduced to 3d and

\begin{align}
\mathcal{A} & =\frac{T_{7}}{2}\sqrt{-\frac{g}{g_{S^{4}}}}e^{-5\phi/4}\frac{R}{r},\nonumber \\
\mathcal{B} & =\frac{T_{7}}{2}\sqrt{-\frac{g}{g_{S^{4}}}}e^{-5\phi/4}\frac{\rho}{R},\nonumber \\
\mathcal{C} & =\frac{T_{7}}{2}\sqrt{-\frac{g}{g_{S^{4}}}}\left[\frac{e^{-5\phi/4}}{R}\left(-\frac{2\zeta}{\rho}+\frac{2\rho}{\zeta}+\frac{3\zeta}{2r}\frac{dr}{d\rho}+\frac{7}{8}\zeta\frac{d\phi}{d\rho}\right)-i\frac{3}{4}\frac{\rho}{R}e^{-\phi/4}\partial_{\zeta}C_{0}\right],\nonumber \\
\mathcal{D} & =\frac{T_{7}}{2}\sqrt{-\frac{g}{g_{S^{4}}}}\left(\frac{1}{2}\frac{e^{-9\phi/4}}{R}+\frac{\rho}{R\zeta}e^{-5\phi/4}\Lambda_{l}^{\pm}\right),\nonumber \\
\Psi & =\frac{1}{\sqrt{\mathcal{A}}}\psi\left(x,\zeta\right)\otimes\varphi^{+l,s}\left(S^{4}\right)\otimes\beta.
\end{align}
The 3+1 dimensional spinor $\psi\left(x,\zeta\right)$ can be further
decomposed by a series of completed functions as,

\begin{equation}
\psi=\sum_{n,m}\left(\begin{array}{c}
\psi_{+}^{\left(n\right)}f_{+}^{\left(n\right)}\\
\psi_{-}^{\left(m\right)}f_{-}^{\left(m\right)}
\end{array}\right).\label{eq:3.13}
\end{equation}
\begin{table}
\begin{centering}
\begin{tabular}{|c|c|c|c|c|c|}
\hline 
$Q=0$ & $n=1$ & $n=2$ & $n=3$ & $n=4$ & $n=5$\tabularnewline
\hline 
\hline 
$M_{\left(n\right)}^{+}$ & 1.65 & 2.89 & 4.18 & 5.61 & 7.26\tabularnewline
\hline 
$M_{\left(n\right)}^{-}$ & 2.56 & 3.55 & 4.87 & 6.32 & 7.94\tabularnewline
\hline 
\end{tabular}
\par\end{centering}
\begin{centering}
\begin{tabular}{|c|c|c|c|c|c|}
\hline 
$Q=1$ & $n=1$ & $n=2$ & $n=3$ & $n=4$ & $n=5$\tabularnewline
\hline 
\hline 
$M_{\left(n\right)}^{+}$ & 1.48 & 2.70 & 4.00 & 5.48 & 7.17\tabularnewline
\hline 
$M_{\left(n\right)}^{-}$ & 2.12 & 3.38 & 4.72 & 6.22 & 7.86\tabularnewline
\hline 
\end{tabular}
\par\end{centering}
\caption{\label{tab:2}Spectrum of the worldvolume fermion in the unit of $M_{KK}=1$
with $\Lambda_{l}=2$.}
\end{table}
Defining the gamma matrix specifically as,

\begin{equation}
\gamma^{\mu}=i\left(\begin{array}{cc}
0 & \sigma^{\mu}\\
\bar{\sigma}^{\mu} & 0
\end{array}\right),\gamma^{4}=\left(\begin{array}{cc}
1 & 0\\
0 & -1
\end{array}\right),
\end{equation}
where $\sigma^{\mu}=\left(1,-\tau^{i}\right),\bar{\sigma}^{\mu}=\left(1,\tau^{i}\right),i=1,2,3$
and $\tau^{i}$ is the Pauli matrix, the action (\ref{eq:3.11}) becomes,

\begin{align}
S_{\mathrm{D7}}= & \sum_{n,m}\int d^{3}xd\zeta\bigg\{-i\psi_{+}^{\left(n\right)\dagger}f_{+}^{\left(n\right)}\bar{\sigma}^{\mu}\partial_{\mu}\psi_{+}^{\left(m\right)}f_{+}^{\left(m\right)}-i\psi_{-}^{\left(m\right)\dagger}f_{-}^{\left(m\right)}\sigma^{\mu}\partial_{\mu}\psi_{-}^{\left(m\right)}f_{-}^{\left(m\right)}\nonumber \\
 & -\psi_{-}^{\left(m\right)\dagger}\psi_{+}^{\left(n\right)}f_{-}^{\left(m\right)}\left[\frac{\mathcal{B}}{\mathcal{A}}\partial_{4}f_{+}^{\left(n\right)}+\left(\frac{\mathcal{C}}{\mathcal{A}}-\frac{\mathcal{B}}{2\mathcal{A}^{2}}\frac{d\mathcal{A}}{d\zeta}\right)f_{+}^{\left(n\right)}+\frac{\mathcal{D}}{\mathcal{A}}f_{+}^{\left(n\right)}\right]\nonumber \\
 & -\psi_{+}^{\left(n\right)\dagger}\psi_{-}^{\left(m\right)}f_{+}^{\left(n\right)}\left[-\frac{\mathcal{B}}{\mathcal{A}}\partial_{4}f_{-}^{\left(m\right)}-\left(\frac{\mathcal{C}}{\mathcal{A}}-\frac{\mathcal{B}}{2\mathcal{A}^{2}}\frac{d\mathcal{A}}{d\zeta}\right)f_{-}^{\left(m\right)}+\frac{\mathcal{D}}{\mathcal{A}}f_{-}^{\left(m\right)}\right]\bigg\},\label{eq:3.15}
\end{align}
which implies the normalization condition for the basis functions
is,

\begin{equation}
\int d\zeta f_{\pm}^{\left(n\right)}f_{\pm}^{\left(m\right)}=\delta^{mn},\label{eq:3.16}
\end{equation}
with eigen equations,

\begin{align}
\frac{\mathcal{B}}{\mathcal{A}}\partial_{4}f_{+}^{\left(n\right)}+\left(\frac{\mathcal{C}}{\mathcal{A}}-\frac{\mathcal{B}}{2\mathcal{A}^{2}}\frac{d\mathcal{A}}{d\zeta}\right)f_{+}^{\left(n\right)}+\frac{\mathcal{D}}{\mathcal{A}}f_{+}^{\left(n\right)} & =M_{\left(n\right)}f_{-}^{\left(n\right)},\nonumber \\
-\frac{\mathcal{B}}{\mathcal{A}}\partial_{4}f_{-}^{\left(n\right)}-\left(\frac{\mathcal{C}}{\mathcal{A}}-\frac{\mathcal{B}}{2\mathcal{A}^{2}}\frac{d\mathcal{A}}{d\zeta}\right)f_{-}^{\left(n\right)}+\frac{\mathcal{D}}{\mathcal{A}}f_{-}^{\left(n\right)} & =M_{\left(n\right)}f_{+}^{\left(n\right)}.\label{eq:3.17}
\end{align}
Taking into account (\ref{eq:3.16}) and (\ref{eq:3.17}), the action
(\ref{eq:3.15}) can be rewritten in a standard kinetic form,

\begin{align}
S_{\mathrm{D7}} & =-\sum_{n}\int d^{3}x\left\{ i\psi_{+}^{\left(n\right)\dagger}\bar{\sigma}^{\mu}\partial_{\mu}\psi_{+}^{\left(n\right)}+i\psi_{-}^{\left(n\right)\dagger}\sigma^{\mu}\partial_{\mu}\psi_{-}^{\left(n\right)}+M_{n}\left[\psi_{-}^{\left(n\right)\dagger}\psi_{+}^{\left(n\right)}+\psi_{+}^{\left(n\right)\dagger}\psi_{-}^{\left(n\right)}\right]\right\} ,\label{eq:3.18}
\end{align}
so the mass spectrum of the worldvolume fermion can be evaluated by
solving (\ref{eq:3.17}) numerically. The resultant mass spectrum
is given in Table \ref{tab:2}, and in particular, its dependence
on the density of instanton $Q$ is illustrated in Figure \ref{fig:2}.
Our numerical calculation indicates the mass spectrum decreases very
slowly at small $Q$ and converges at large $Q$.

In this work, since the concerned fermionic flux $\psi$ on the D7-brane
is the baryonic field produced by the $N_{c}$ open strings, an overall
factor for its action (\ref{eq:3.18}) must be picked up to include
the $N_{c}$ contributions of the $N_{c}$ open strings. The simple
way to achieve this is to impose the rescaling for quarks in the large
$N_{c}$ QCD as $\psi_{\pm}^{\left(n\right)}\rightarrow\sqrt{N_{c}}\psi_{\pm}^{\left(n\right)}$
to (\ref{eq:3.18}) because the $N_{c}$ open strings are identified
to quarks in the boundary theory. Another reason we can use this rescaling
is that a hadron in the large $N_{c}$ limit behaves like a classical
particle \cite{key-40}. Therefore, we can define the fixed part of
the baryonic field as \cite{key-40} in the large $N_{c}$ limit,
which means the exact quadratic form for the baryonic action should
be,

\begin{equation}
S_{\mathrm{D7}}=-N_{c}\sum_{n}\int d^{3}x\left\{ i\psi_{+}^{\left(n\right)\dagger}\bar{\sigma}^{\mu}\partial_{\mu}\psi_{+}^{\left(n\right)}+i\psi_{-}^{\left(n\right)\dagger}\sigma^{\mu}\partial_{\mu}\psi_{-}^{\left(n\right)}+M_{n}\left[\psi_{-}^{\left(n\right)\dagger}\psi_{+}^{\left(n\right)}+\psi_{+}^{\left(n\right)\dagger}\psi_{-}^{\left(n\right)}\right]\right\} .\label{eq:3.19}
\end{equation}
Thus, the Hamiltonian associated to (\ref{eq:3.19}) can be written
in the momentum space by using the Fourier transformation $\psi\sim a_{\mathbf{p}}^{s}u_{\mathbf{p}}^{s}e^{-iE_{\mathbf{p}}t+\mathbf{p}\cdot\mathbf{x}}+b_{\mathbf{p}}^{s\dagger}v_{\mathbf{p}}^{s}e^{iE_{\mathbf{p}}t-\mathbf{p}\cdot\mathbf{x}}$
with the creation/annihilation operators as,

\begin{equation}
H=N_{c}\int\frac{d^{3}\mathbf{p}}{\left(2\pi\right)^{3}}\sum_{s=+,-}E_{\mathbf{p}}\left(a_{\mathbf{p}}^{s\dagger}a_{\mathbf{p}}^{s}+b_{\mathbf{p}}^{s\dagger}b_{\mathbf{p}}^{s}\right),
\end{equation}
where $E_{\mathbf{p}}$ is the unit energy eigenvalue as $E_{\mathbf{p}}=\sqrt{\mathbf{p}^{2}+M_{n}^{2}}$
and $M_{n}$ is one of the given mass in the fermionic mass spectrum.
Therefore we can see the average value of the Hamiltonian $H$ on
a single baryon state $\left|1\right\rangle =a_{\mathbf{p}}^{s\dagger}\left|0\right\rangle $
is $\left\langle 1\left|H\right|1\right\rangle =N_{c}E_{\mathbf{p}}$
, and in particular $\left\langle 1\left|H\right|1\right\rangle =N_{c}M_{n}$
for $\mathbf{p}=0$ which means the baryon mass is proportional to
$N_{c}$ as it is expected for a baryon in the large $N_{c}$ limit.
Moreover, although this model is 3d, it is interesting to compare
our numerical results with the experimental data of $N_{c}=3$ baryon.
The quantum numbers $l,s$ presented in (\ref{eq:3.8}) can be identified
to angular momentum and isospin for a baryon, and ``$\pm$'' refers
to the chirality of the fermion under $\gamma^{4}$ which can be naturally
identified to the parity of a baryon. Thus the lowest fermion in our
model takes $l=0,s=\pm\frac{1}{2}$ and even parity with mass $M_{\left(1\right)}^{+}\simeq1.65M_{KK}N_{c}\simeq4.95M_{KK}$
according to Table \ref{tab:2}. In the experimental data of baryon,
such a baryon state is recognized to be proton or neutron. On the
other hand, the lowest vector takes the mass as $m_{1}\simeq3.15M_{KK}$
according to Table \ref{tab:A1} and in the experimental data of meson,
such a vector meson is recognized to be $\rho$ meson. So the mass
ratio of proton (or neutron) and $\rho$ meson in our model is $M_{\left(1\right)}^{+}/m_{1}\simeq1.57$
which is close to the experimental data $M_{\mathrm{proton}}/M_{\rho\mathrm{-meson}}\simeq1.22$. 

\subsection{The spectral function}

In this section, let us investigate the fermionic spectrum in a parallel
way, that is to evaluate the correlation function as its spectral
function with the holographic prescription. Note that the holographic
prescription for computing the two-point correlation function is also
valid with respect to the deconfined background. In this section,
we first review the holographic prescription for computing the two-point
correlation function with a spinor, then use it to analyze the spectral
function both in confined and deconfined geometry.

\subsubsection*{The holographic prescription}

Let us recall the principle of the AdS/CFT for computing the two-point
correlation function with spinor \cite{key-41,key-42}. First, the
generating functional of the dual quantum field theory (QFT) $Z_{\mathrm{QFT}}$
is equal to its gravitational partition function $Z_{\mathrm{gravity}}$
in the bulk geometry. Therefore, for a spinor field $\psi$ in the
bulk, we have

\begin{equation}
Z_{\mathrm{QFT}}\left[\bar{\psi}_{0},\psi_{0}\right]=Z_{\mathrm{gravity}}\left[\bar{\psi},\psi\right]\big|_{\bar{\psi},\psi\rightarrow\bar{\psi}_{0},\psi_{0}},\label{eq:3.21}
\end{equation}
with

\begin{align}
Z_{\mathrm{QFT}}\left[\bar{\psi_{0}},\psi_{0}\right] & =\left\langle \exp\left\{ \int_{\partial\mathcal{M}}\left(\bar{\eta}\psi_{0}+\bar{\psi}_{0}\eta\right)d^{D}x\right\} \right\rangle ,\nonumber \\
Z_{\mathrm{gravity}}\left[\bar{\psi},\psi\right] & =e^{-S_{\mathrm{gravity}}^{ren}},\nonumber \\
S_{\mathrm{gravity}}^{ren} & =\int_{\mathcal{M}}\mathcal{L}_{\mathrm{gravity}}^{ren}\left[\bar{\psi},\psi\right]\sqrt{-g}d^{D+1}x
\end{align}
Here $\mathcal{M}$ denotes the $D+1$ dimensional bulk space and
its holographic boundary is given at $\partial\mathcal{M}$. $\psi_{0}$
refers to the boundary value of $\psi$ as a source of the boundary
fermionic operator $\eta$, $\mathcal{L}_{\mathrm{gravity}}^{ren}$
represents the renormalized Lagrangian of the bulk field $\psi$.
Then follow the standard steps in quantum field theory (QFT), the
one-point function i.e. average value of $\bar{\eta}$ is given as

\begin{equation}
\left\langle \bar{\eta}\right\rangle =\frac{1}{Z_{\mathrm{QFT}}}\frac{\delta Z_{\mathrm{QFT}}}{\delta\psi_{0}}=\frac{1}{Z_{\mathrm{gravity}}}\frac{\delta Z_{\mathrm{gravity}}}{\delta\psi_{0}}=-\frac{\delta S_{\mathrm{gravity}}^{ren}}{\delta\psi_{0}}\equiv\Pi_{0},\label{eq:3.23}
\end{equation}
where we have used the relation (\ref{eq:3.21}). Therefore, the two-point
correlation function $G_{R}$ of $\eta$ is obtained as,

\begin{equation}
\left\langle \bar{\eta}\left(\omega,\vec{k}\right)\right\rangle =G_{R}\left(\omega,\vec{k}\right)\psi_{0},
\end{equation}
i.e.

\begin{equation}
\Pi_{0}=G_{R}\left(\omega,\vec{k}\right)\psi_{0},
\end{equation}
where $\omega,\vec{k}$ refers to the frequency and spacial momentum
of the associated Fourier modes. Altogether, it is possible to evaluate
$\psi_{0},\Pi_{0}$ in order to evaluate the two-point correlation
function $G_{R}$ of $\eta$ by solving the classical gravity action
$S_{\mathrm{gravity}}^{ren}$ in holography. We note that the spinor
$\eta,\psi_{0}$ must be gauge-invariant operator under the color
group $U\left(N_{c}\right)$ i.e. they should be baryonic field\footnote{The $N_{c}$-counting for the two-point correlation function implies
an overall factor $N_{c}^{-1}$ should be picked up in front of $G_{R}$
\cite{key-40} if we use the rescaling $\psi\rightarrow\sqrt{N_{c}}\psi$.}.

\subsubsection*{Approach to the confined phase}

\begin{figure}
\begin{centering}
\includegraphics[scale=0.27]{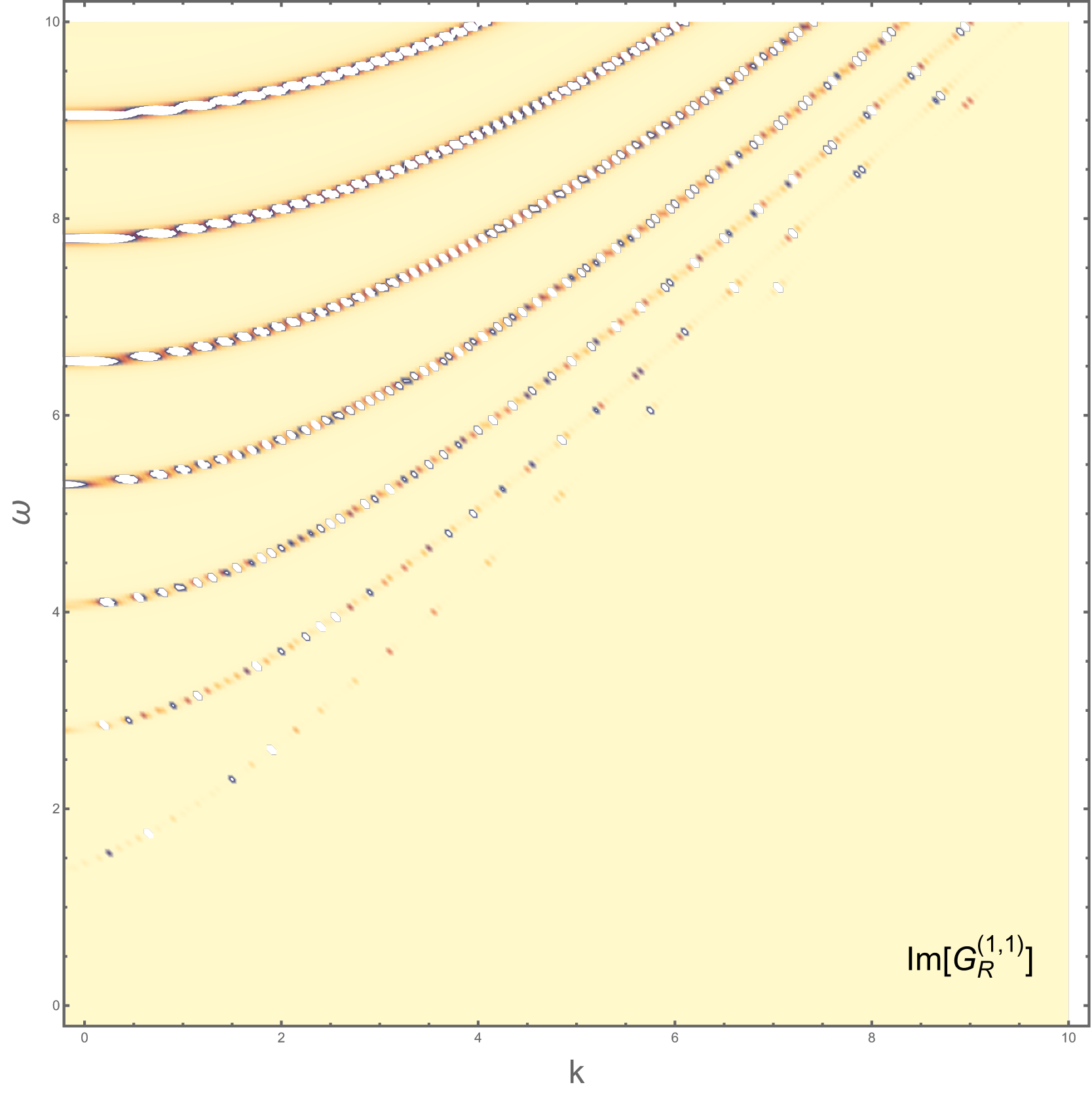}\includegraphics[scale=0.27]{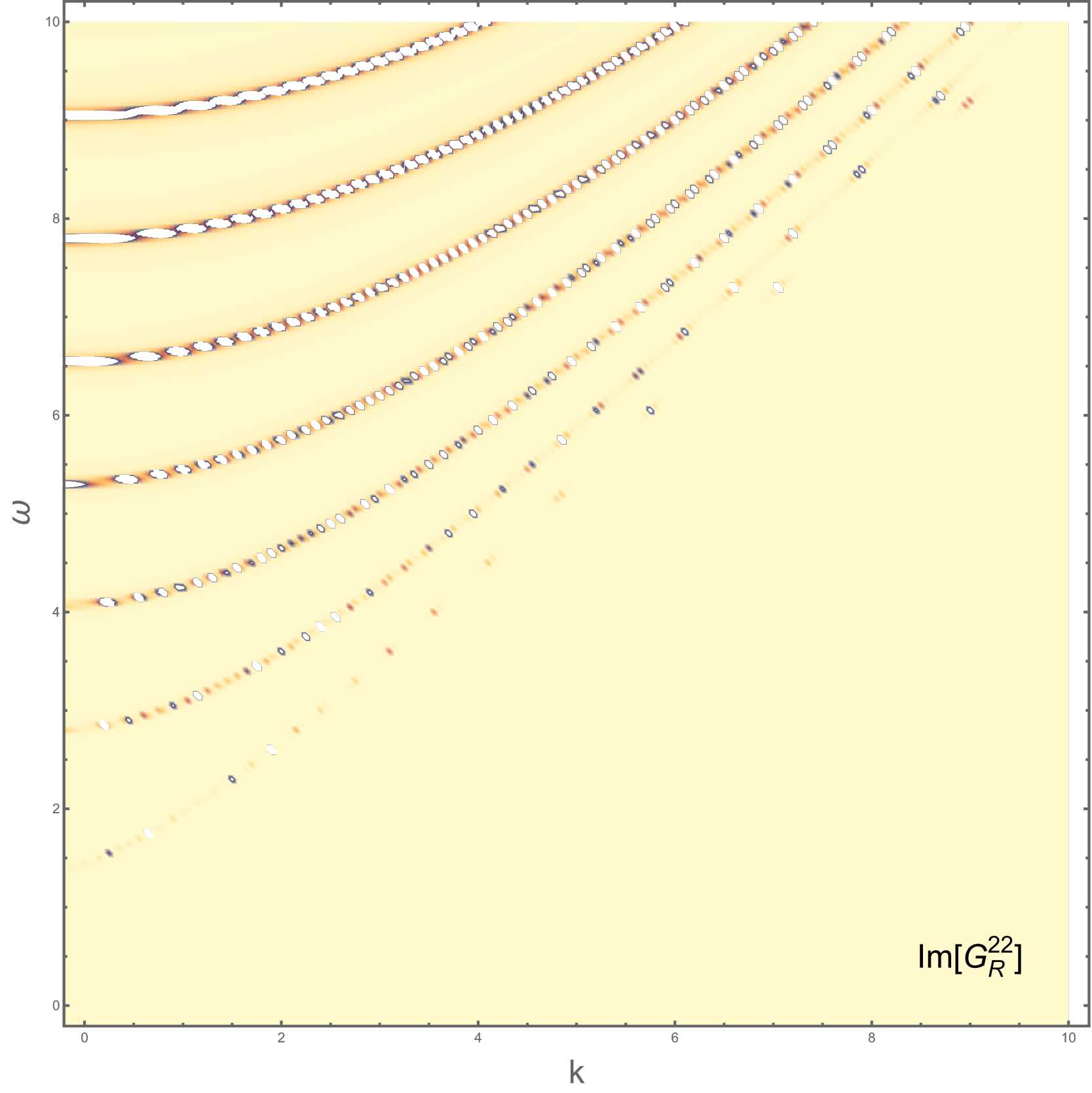}
\par\end{centering}
\caption{\label{fig:3}The confined dispersion curves from the fermionic Green
function $G_{R}^{\left(1,1\right)}$ and $G_{R}^{\left(2,2\right)}$
as functions of $\omega$ and $\mathrm{k}$ in the unit of $M_{KK}=1$.
The red-white points denote the peaks in the correlation function.}
\end{figure}

\begin{figure}
\begin{centering}
\includegraphics[scale=0.27]{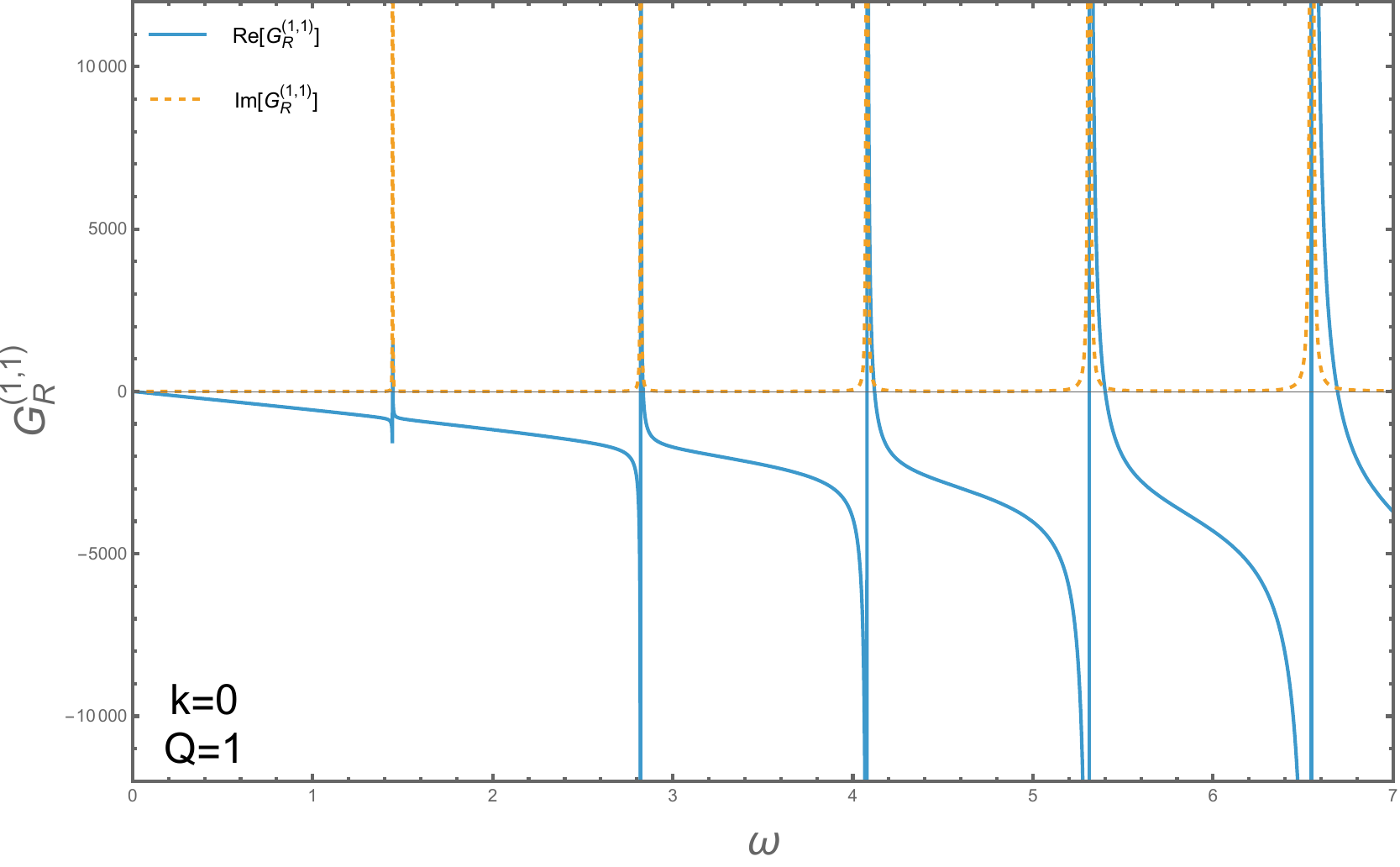}\includegraphics[scale=0.27]{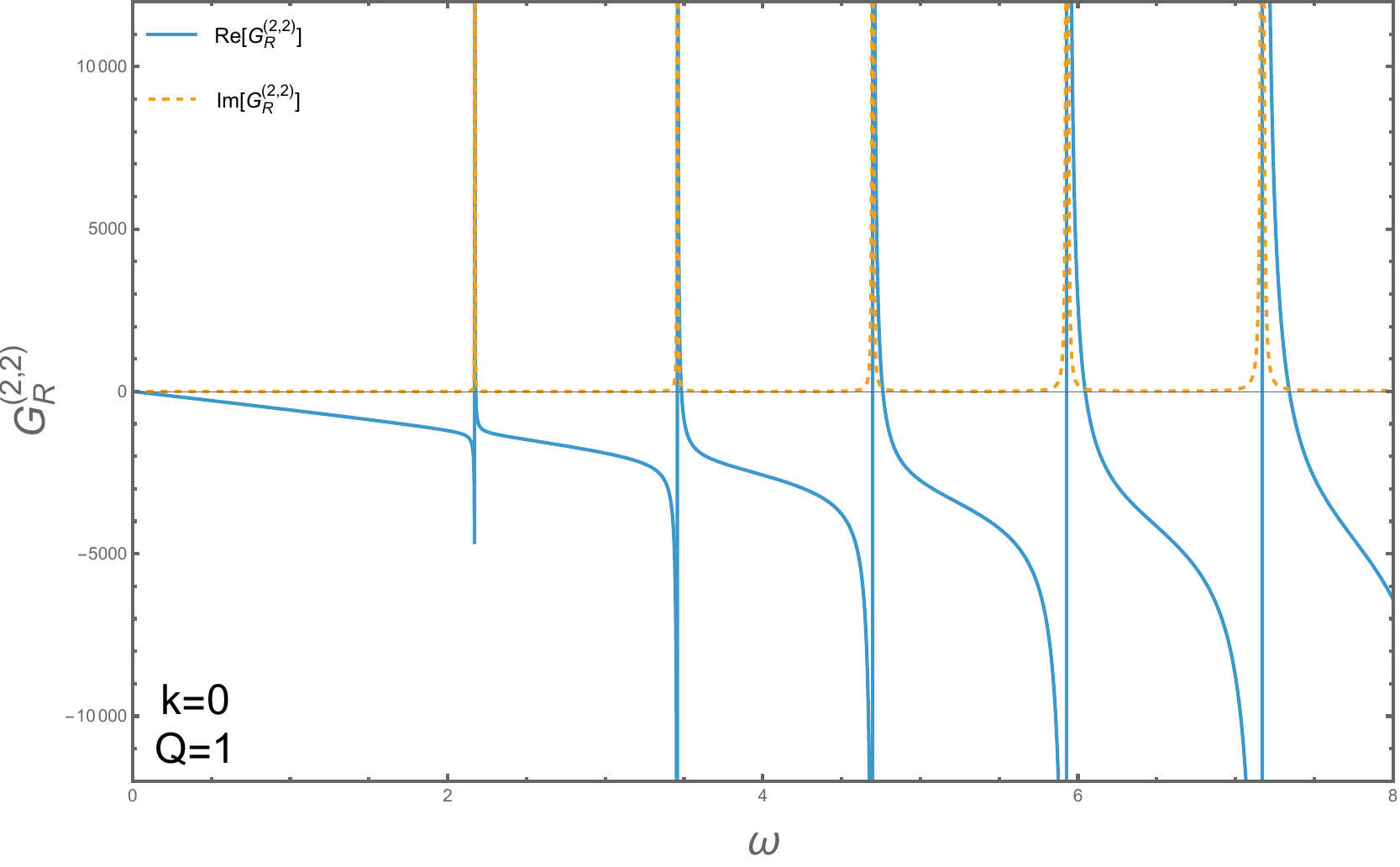}
\par\end{centering}
\begin{centering}
\includegraphics[scale=0.27]{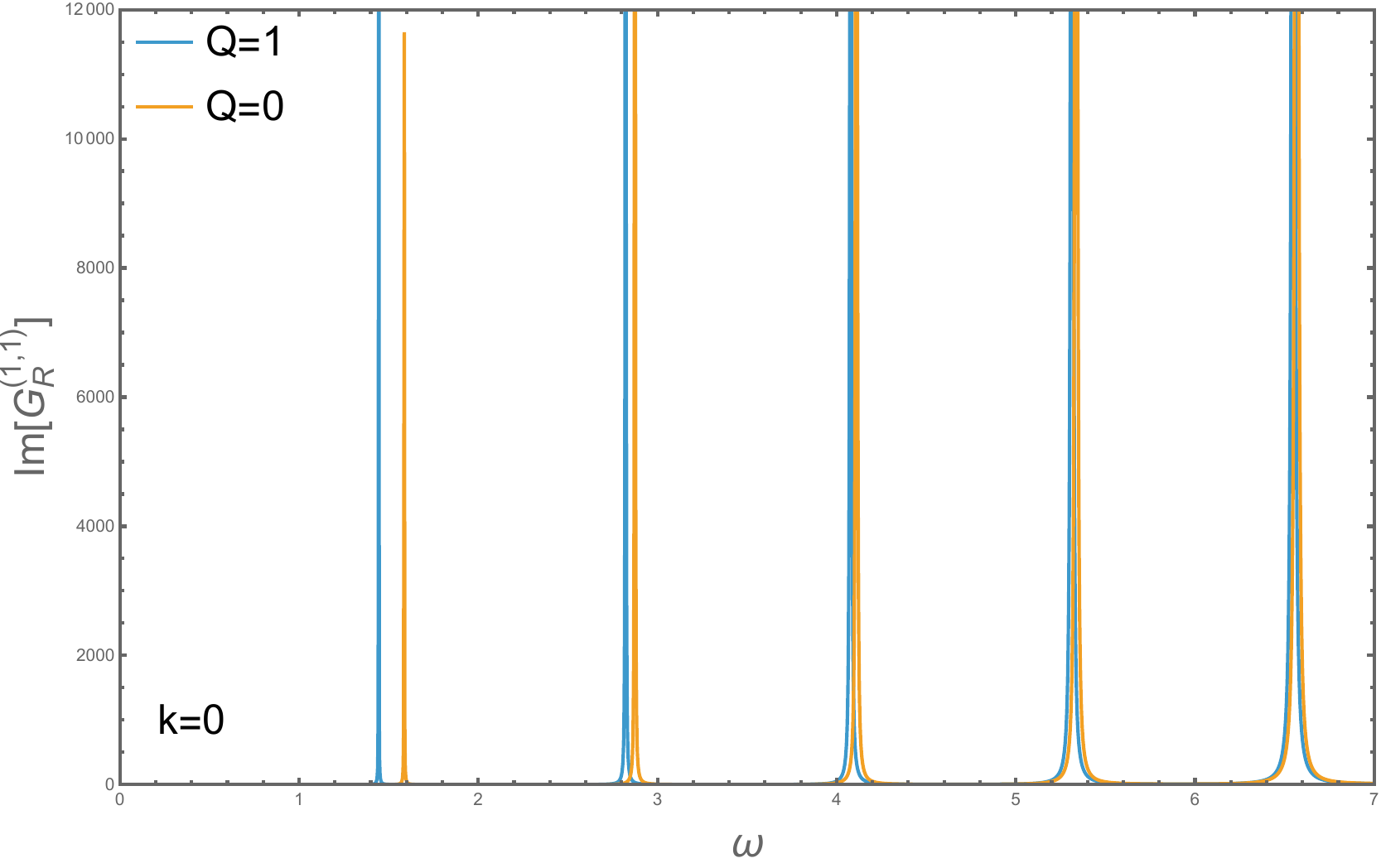}\includegraphics[scale=0.27]{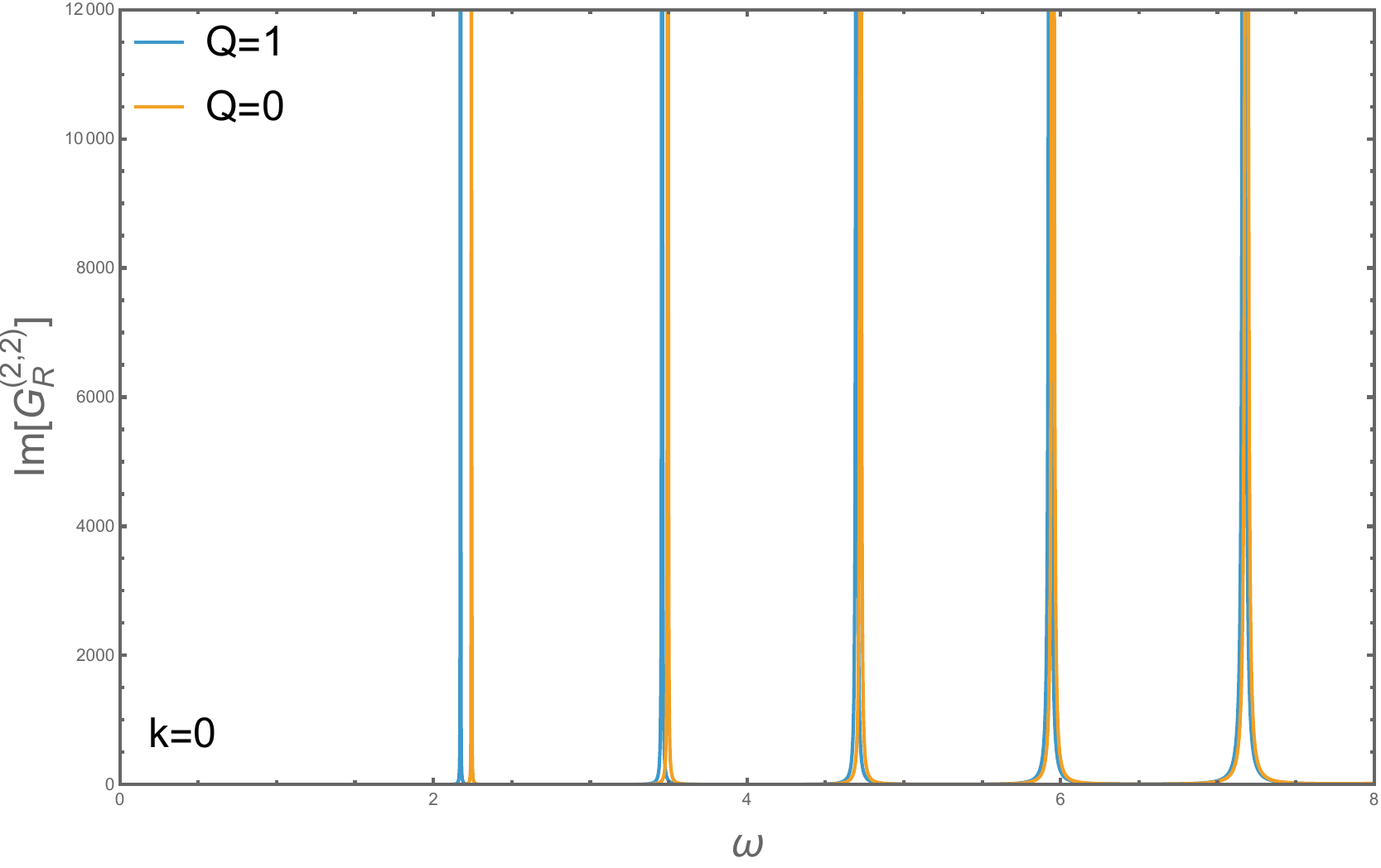}
\par\end{centering}
\caption{\label{fig:4}The confined fermionic Green function $G_{R}^{\left(1,1\right)}$
and $G_{R}^{\left(2,2\right)}$ as functions of $\omega$ with $\mathrm{k}=0$
and various $Q$ in the unit of $M_{KK}=1$.}
\end{figure}

Let us approach the above holographic prescription with the confined
metric given in (\ref{eq:2.11}). The action for the bulk field $\psi$
on the worldvolume of the D7-brane takes the same form as (\ref{eq:3.5})
which can be reduced to an effectively 3d form as it is given in (\ref{eq:3.11}).
So its boundary term corresponds to

\begin{equation}
S_{\mathrm{bdry}}=i\int d^{3}xd\Omega_{4}\left[\bar{\Psi}\mathcal{B}\gamma^{4}\Psi\big|_{\partial\mathcal{M}}+...\right].
\end{equation}
To obtain a standard boundary form, we redefine the spinor $\psi$
by rescaling $\Psi\rightarrow\sqrt{\mathcal{B}}\Psi$ , then the action
(\ref{eq:2.11}) can be written as,

\begin{equation}
S_{\mathrm{D7}}=i\int d^{3}xd\zeta\left(\bar{\psi}\frac{\mathcal{A}}{\mathcal{B}}\gamma^{\mu}\partial_{\mu}\psi+\bar{\psi}\gamma^{4}\partial_{4}\psi-\frac{1}{2}\bar{\psi}\gamma^{4}\psi\partial_{4}\ln\mathcal{B}+\bar{\psi}\frac{\mathcal{C}}{\mathcal{B}}\gamma^{4}\psi+\bar{\psi}\frac{\mathcal{D}}{\mathcal{B}}\psi\right),
\end{equation}
and the Dirac equation is derived as,

\begin{equation}
\left[\frac{\mathcal{A}}{\mathcal{B}}\gamma^{\mu}\partial_{\mu}+\gamma^{4}\partial_{4}+\left(\frac{\mathcal{C}}{\mathcal{B}}-\frac{1}{2}\partial_{4}\ln\mathcal{B}\right)\gamma^{4}+\frac{\mathcal{D}}{\mathcal{B}}\right]\psi=0.\label{eq:3.28}
\end{equation}
Insert the Fourier modes as the ansatz,

\begin{equation}
\psi=\int\frac{d^{4}p}{\left(2\pi\right)^{4}}e^{ik\cdot x}\chi\left(\zeta,k\right)=\int\frac{d^{4}p}{\left(2\pi\right)^{4}}e^{ik\cdot x}\left[\begin{array}{c}
\chi_{+}\left(\zeta,k\right)\\
\chi_{-}\left(\zeta,k\right)
\end{array}\right],\label{eq:3.29}
\end{equation}
 into (\ref{eq:3.28}), the Dirac equation (\ref{eq:3.28}) becomes,

\begin{align}
\left(\partial_{4}+\frac{\mathcal{C}}{\mathcal{B}}-\frac{1}{2}\partial_{4}\ln\mathcal{B}+\frac{\mathcal{D}}{\mathcal{B}}\right)\chi_{+}+\frac{\mathcal{A}}{\mathcal{B}}\left(\tau^{3}k+\omega\right)\chi_{-} & =0\nonumber \\
\frac{\mathcal{A}}{\mathcal{B}}\left(-\tau^{3}k+\omega\right)\chi_{+}+\left(-\partial_{4}-\frac{\mathcal{C}}{\mathcal{B}}+\frac{1}{2}\partial_{4}\ln\mathcal{B}+\frac{\mathcal{D}}{\mathcal{B}}\right)\chi_{-} & =0,\label{eq:3.30}
\end{align}
where the momentum is set to be along $x^{2}$ direction as, $k_{\mu}=\left(-\omega,0,\mathrm{k}\right)$
and the gamma matrices are chosen as,

\begin{equation}
\gamma^{0}=i\left(\begin{array}{cc}
0 & 1\\
1 & 0
\end{array}\right),\gamma^{2}=i\left(\begin{array}{cc}
0 & -\tau^{3}\\
\tau^{3} & 0
\end{array}\right),\gamma^{4}=\left(\begin{array}{cc}
1 & 0\\
0 & -1
\end{array}\right).\label{eq:3.31}
\end{equation}
The equation (\ref{eq:3.30}) reduces to two second order differential
equation for $\chi_{+}$ and $\chi_{-}$ respectively which can be
solved analytically near the boundary $\zeta\rightarrow\infty$, as

\begin{align}
\chi_{+} & =R_{A}\zeta^{\Lambda_{l}^{+}-\frac{3}{2}}+R_{B}\zeta^{-\Lambda_{l}^{+}-\frac{3}{2}},\nonumber \\
\chi_{-} & =L_{A}\zeta^{\Lambda_{l}^{+}-\frac{1}{2}}+L_{B}\zeta^{-\Lambda_{l}^{+}-\frac{5}{2}},\label{eq:3.32}
\end{align}
where $L_{A,B},R_{A,B}$ are constant spinors. Therefore, the boundary
value of $\psi$ can be defined by taking the dominated term in (\ref{eq:3.32})
as,

\begin{equation}
\psi|_{\partial\mathcal{M}}=\lim_{\zeta\rightarrow\infty}\zeta^{-\Lambda_{l}^{+}+\frac{1}{2}}\psi=\left(\begin{array}{c}
0\\
L_{A}
\end{array}\right),\label{eq:3.33}
\end{equation}
so the $\Pi_{0}$ given in (\ref{eq:3.23}) becomes,

\begin{equation}
\Pi_{0}|_{\partial\mathcal{M}}\sim-\lim_{\zeta\rightarrow\infty}\zeta^{-\Lambda_{l}^{+}+\frac{3}{2}}\chi_{+}|_{\partial\mathcal{M}}=-R_{A}.\label{eq:3.34}
\end{equation}
Altogether, the renormalized two-point Green function can be obtained
as

\begin{equation}
G_{R}^{\left(h,h\right)}=\left(-1\right)^{h}\lim_{\zeta\rightarrow\infty}\zeta\frac{\chi_{+}^{\left(h\right)}}{\chi_{-}^{\left(h\right)}},h=1,2\label{eq:3.35}
\end{equation}
By introducing ratios

\begin{equation}
\xi_{\left(h\right)}=\left(-1\right)^{h+1}\frac{\chi_{+}^{\left(h\right)}}{\chi_{-}^{\left(h\right)}},\chi_{\pm}=\left[\begin{array}{c}
\chi_{\pm}^{\left(1\right)}\\
\chi_{\pm}^{\left(2\right)}
\end{array}\right],\label{eq:3.36}
\end{equation}
the equations of motion for $\xi_{\left(h\right)}$ can be obtained
from the Dirac equation as,

\begin{align}
\frac{\mathcal{B}}{\mathcal{A}}\xi_{\left(1\right)}^{\prime} & =-k+\frac{2\mathcal{D}}{\mathcal{A}}\xi_{\left(1\right)}+k\xi_{\left(1\right)}^{2}+\omega+\omega\xi_{\left(1\right)}^{2},\nonumber \\
\frac{\mathcal{B}}{\mathcal{A}}\xi_{\left(2\right)}^{\prime} & =-k+\frac{2\mathcal{D}}{\mathcal{A}}\xi_{\left(2\right)}+k\xi_{\left(2\right)}^{2}-\omega-\omega\xi_{\left(2\right)}^{2},\label{eq:3.37}
\end{align}
where ``$\prime$'' refers to the derivative with respect to $\zeta$.
The equations in (\ref{eq:3.37}) can be solved numerically with the
Wick-rotated version of the infalling boundary condition as $\xi_{\left(h\right)}|_{r=r_{KK}}=\left(-1\right)^{h}$,
since the confined metric (\ref{eq:2.11}) can be obtained by a double
Wick rotation from the deconfined metric (\ref{eq:2.10}). With all
above, the numerical results for the Green functions are given in
Figure \ref{fig:3} and \ref{fig:4}. 
 
As the two-point Green function is proportional to $\left(\cancel{k}-M_{n}\right)^{-1}$
, the location of the peaks in the Green function refers to the onset
mass of the fermion which agrees with the mass spectrum we obtained
in Section 3.2. In addition, Figure \ref{fig:4} also displays that
the onset mass of the fermion decreases very slowly when the instanton
density $Q$ increases. This conclusion agrees well with the behavior
of the mass spectrum given in Figure \ref{fig:2}.

\subsubsection*{Approach to the deconfined phase}

\begin{figure}
\begin{centering}
\includegraphics[scale=0.27]{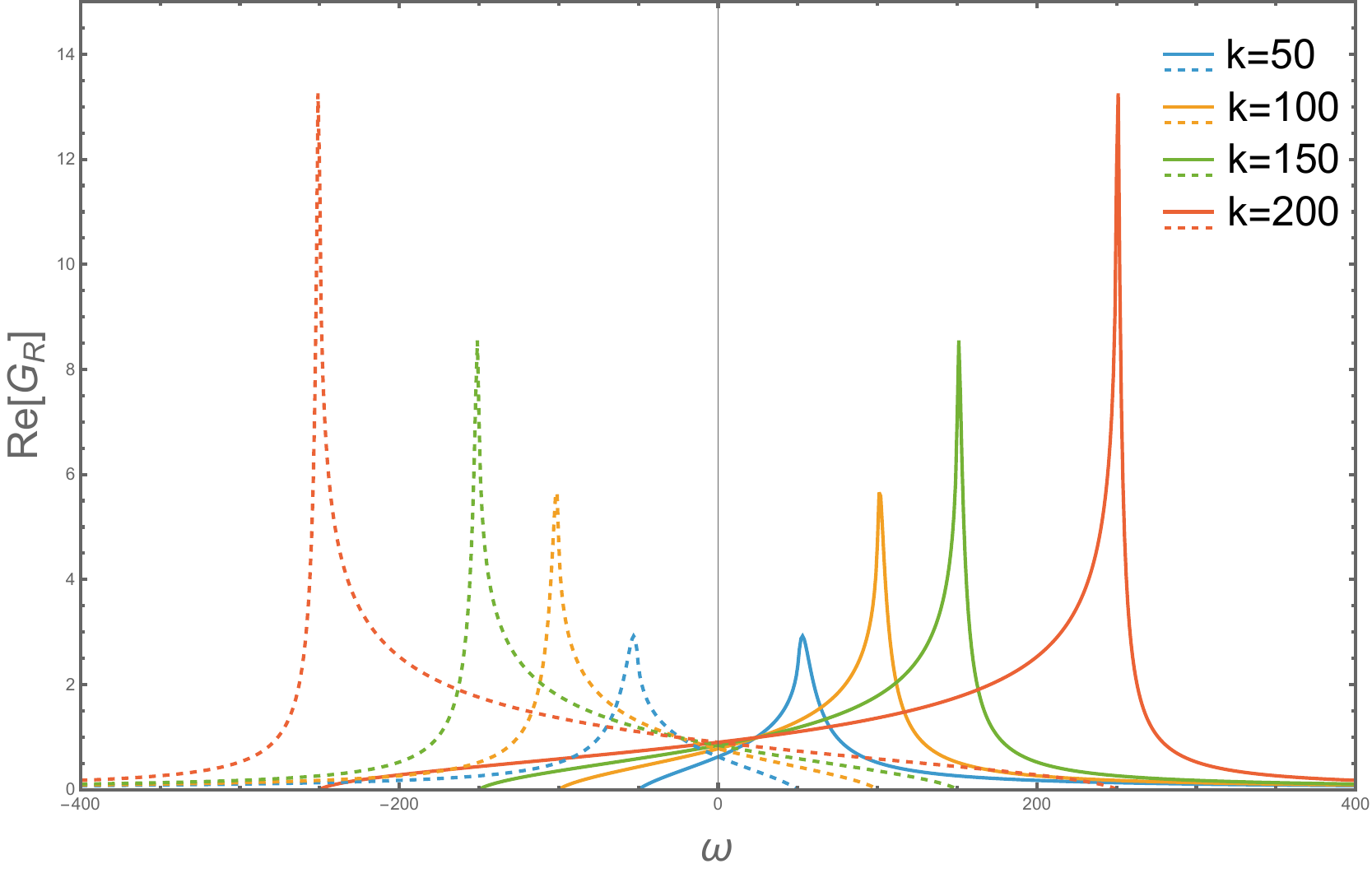}\includegraphics[scale=0.27]{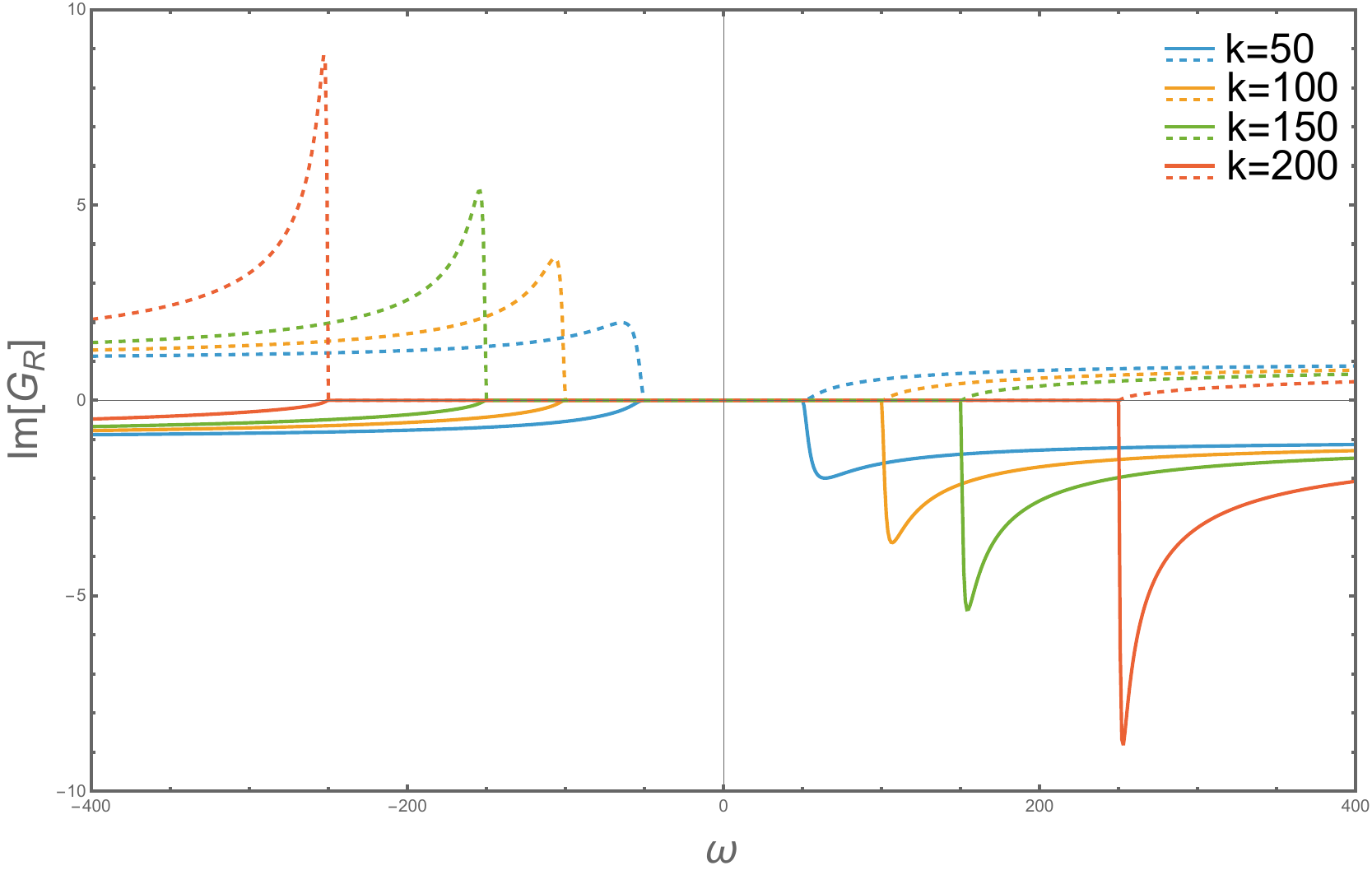}
\par\end{centering}
\caption{\label{fig:5}The deconfined fermionic Green function from D3/D7 model
with instanton density $Q=1$ and temperature $T=\frac{1}{\pi}$.
The solid and dashed lines refer respectively to $G_{R}^{\left(1,1\right)}$
and $G_{R}^{\left(2,2\right)}$ with various momentums $\mathrm{k}$.}
\end{figure}

\begin{figure}
\begin{centering}
\includegraphics[scale=0.2]{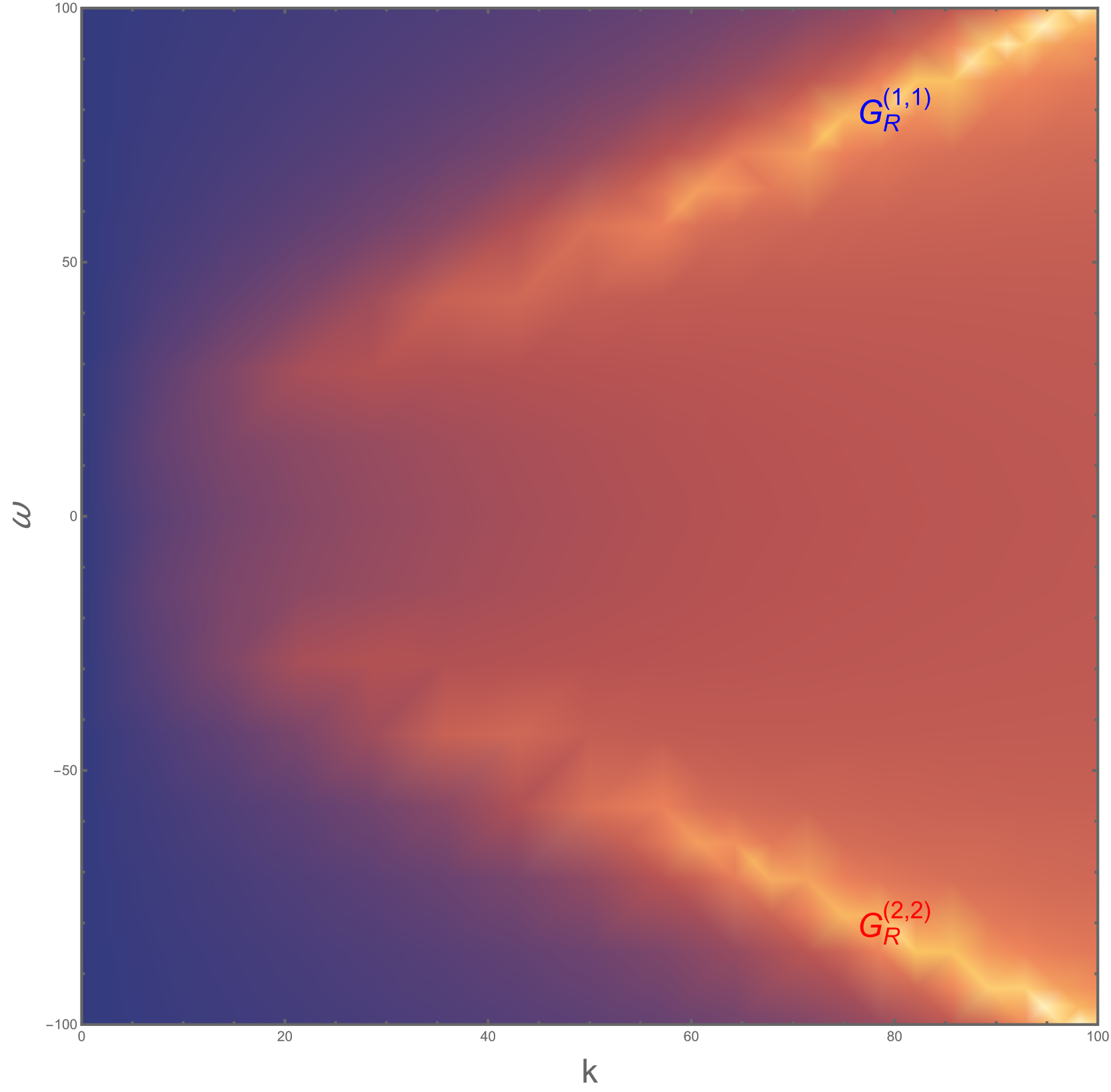}\includegraphics[scale=0.31]{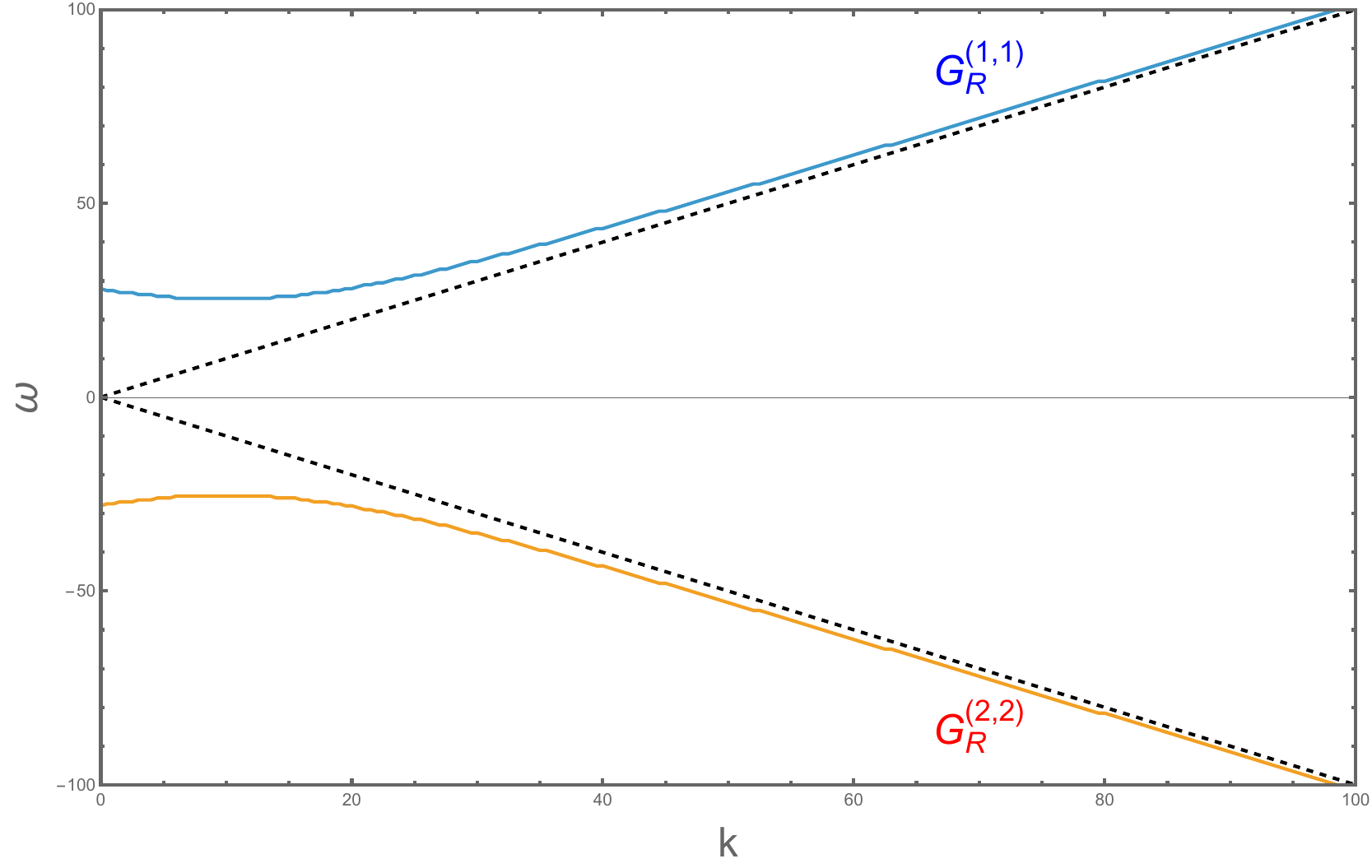}
\par\end{centering}
\caption{\label{fig:6}The dispersion curves from the fermionic Green function
with $Q=1$ and $T=\frac{1}{\pi}$. \textbf{Left:} The density plot
of the fermionic Green functions where the white region refers to
the peaks in the Green function. \textbf{Right:} The numerical evaluation
of the dispersion curves from fermionic Green functions where the
dashed lines refer to the light cone.}

\end{figure}

In the deconfined geometry, there is not warp configuration for a
D5-brane on $S^{5}$ in the background (\ref{eq:2.2}) \cite{key-a5},
therefore, as most works about gauge-gravity duality, the bulk fermion
$\psi$ is identified to baryonic plasmino instead of baryon \cite{key-41,key-42,key-44,key-45}.
With this in hand, let us follow the above steps with respect to the
deconfined geometry given in (\ref{eq:2.10}) in order to compute
the correlation function. To begin with, the Dirac operator is computed
as
\begin{align}
\Gamma^{\alpha}\nabla_{\alpha}= & \frac{e^{-\frac{\phi}{4}}}{\sqrt{f}}\frac{R}{r}\gamma^{0}\partial_{0}+e^{-\frac{\phi}{4}}\frac{R}{r}\gamma^{i}\partial_{i}+e^{-\frac{\phi}{4}}\frac{\rho}{R}\gamma^{4}\partial_{4}\nonumber \\
 & +\frac{e^{-\frac{\phi}{4}}}{R}\gamma^{4}\left(-\frac{2\zeta}{\rho}+\frac{2\rho}{\zeta}+\frac{\zeta}{4f}\frac{df}{d\rho}+\frac{3\zeta}{2r}\frac{dr}{d\rho}+\frac{7\zeta}{8}\frac{d\phi}{d\rho}\right)+e^{-\phi/4}\frac{\rho}{R\zeta}\cancel{D}_{S^{4}}.\label{eq:3.38}
\end{align}
Put (\ref{eq:3.38}) with (\ref{eq:3.7}) into the fermionic action
(\ref{eq:3.5}), we can obtain,

\begin{align}
S_{\mathrm{D7}}^{f}= & i\int d^{3}xd\zeta\bar{\psi}\left[\frac{\mathcal{A}}{\mathcal{C}}\gamma^{0}\partial_{0}+\frac{\mathcal{B}}{\mathcal{C}}\gamma^{i}\partial_{i}+\gamma^{4}\left(\partial_{4}-\frac{1}{2}\partial_{4}\ln\mathcal{C}\right)+\frac{\mathcal{D}}{\mathcal{C}}\gamma^{4}+\frac{\mathcal{E}}{\mathcal{C}}\right]\psi,\label{eq:3.39}
\end{align}
by using the dimensional reduction given in Section 3.2. And the corresponding
coefficient terms are collected as,

\begin{align}
\mathcal{A} & =\frac{T_{7}}{2}e^{-5\phi/4}\sqrt{-\frac{g}{g_{S^{4}}}}\frac{R}{r\sqrt{f}},\nonumber \\
\mathcal{B} & =\frac{T_{7}}{2}e^{-5\phi/4}\sqrt{-\frac{g}{g_{S^{4}}}}\frac{R}{r},\nonumber \\
\mathcal{C} & =\frac{T_{7}}{2}e^{-5\phi/4}\sqrt{-\frac{g}{g_{S^{4}}}}\frac{\rho}{R},\nonumber \\
\mathcal{D} & =\frac{T_{7}}{2}\sqrt{-\frac{g}{g_{S^{4}}}}\left[\frac{e^{-5\phi/4}}{R}\left(-\frac{2\zeta}{\rho}+\frac{2\rho}{\zeta}+\frac{\zeta}{4f}\frac{df}{d\rho}+\frac{3\zeta}{2r}\frac{dr}{d\rho}+\frac{7\zeta}{8}\frac{d\phi}{d\rho}\right)-i\frac{3}{4}\frac{\rho}{R}e^{\phi/4}\partial_{\zeta}C_{0}\right],\nonumber \\
\mathcal{E} & =\frac{T_{7}}{2}\sqrt{-\frac{g}{g_{S^{4}}}}\left(\frac{1}{2}\frac{e^{-9\phi/4}}{R}+e^{-5\phi/4}\frac{\rho}{R\zeta}\Lambda_{l}^{+}\right),\nonumber \\
\Psi & =\frac{1}{\sqrt{\mathcal{C}}}\psi\left(x,\zeta\right)\otimes\varphi^{+l,s}\left(S^{4}\right)\otimes\beta.
\end{align}
Then the Dirac equation derived from (\ref{eq:3.39}) is,

\begin{equation}
\left[\frac{\mathcal{A}}{\mathcal{C}}\gamma^{0}\partial_{0}+\frac{\mathcal{B}}{\mathcal{C}}\gamma^{i}\partial_{i}+\gamma^{4}\left(\partial_{4}-\frac{1}{2}\partial_{4}\ln\mathcal{C}\right)+\frac{\mathcal{D}}{\mathcal{C}}\gamma^{4}+\frac{\mathcal{E}}{\mathcal{C}}\right]\psi=0.\label{eq:3.41}
\end{equation}
Recall the ansatz (\ref{eq:3.29}) and gamma matrices given in (\ref{eq:3.31}),
the Dirac equation becomes,

\begin{align}
\left(\partial_{4}-\frac{1}{2}\partial_{4}\ln\mathcal{C}+\frac{\mathcal{D}}{\mathcal{C}}+\frac{\mathcal{E}}{\mathcal{C}}\right)\chi_{+}+\left(\omega\frac{\mathcal{A}}{\mathcal{C}}+\tau^{3}\mathrm{k}\frac{\mathcal{B}}{\mathcal{C}}\right)\chi_{-} & =0,\nonumber \\
\left(\omega\frac{\mathcal{A}}{\mathcal{C}}-\tau^{3}\mathrm{k}\frac{\mathcal{B}}{\mathcal{C}}\right)\chi_{+}+\left(-\partial_{4}+\frac{1}{2}\partial_{4}\ln\mathcal{C}-\frac{\mathcal{D}}{\mathcal{C}}+\frac{\mathcal{E}}{\mathcal{C}}\right)\chi_{-} & =0,\label{eq:3.42}
\end{align}
which can be solved analytically near the holographic boundary $\zeta\rightarrow\infty$,
as

\begin{align}
\chi_{+} & =R_{A}\zeta^{\Lambda_{l}^{+}+1}+R_{B}\zeta^{-\Lambda_{l}^{+}-1},\nonumber \\
\chi_{-} & =L_{A}\zeta^{\Lambda_{l}^{+}}+L_{B}\zeta^{-\Lambda_{l}^{+}}.
\end{align}
Follow the same steps given in (\ref{eq:3.33}) and (\ref{eq:3.34}),
the Green function remains to be (\ref{eq:3.35}). So by introducing
ratios given in (\ref{eq:3.36}), the equations of motion for $\xi_{\left(s\right)}$
can be obtained from the Dirac equation (\ref{eq:3.42}) as,

\begin{align}
\frac{\mathcal{C}}{\mathcal{B}}\xi_{\left(1\right)}^{\prime} & =-\mathrm{k}-\frac{\mathcal{A}}{\mathcal{B}}\omega-\frac{2\mathcal{E}}{\mathcal{B}}\xi_{\left(1\right)}+\left(\mathrm{k}-\frac{\mathcal{A}}{\mathcal{B}}\omega\right)\xi_{\left(1\right)}^{2},\nonumber \\
\frac{\mathcal{C}}{\mathcal{B}}\xi_{\left(2\right)}^{\prime} & =-\mathrm{k}+\frac{\mathcal{A}}{\mathcal{B}}\omega-2\frac{\mathcal{E}}{\mathcal{B}}\xi_{\left(2\right)}+\left(\mathrm{k}+\frac{\mathcal{A}}{\mathcal{B}}\omega\right)\xi_{\left(2\right)}^{2},\label{eq:3.44}
\end{align}
which can be solved numerically with the infalling boundary condition
$\xi_{\left(h\right)}|_{r=r_{H}}=\left(-1\right)^{h}i$. The numerical
results are collected in Figure \ref{fig:5} and \ref{fig:6}. 
 
It indicates that dispersion curves satisfy $\omega\simeq\mathrm{k}$
for large $\mathrm{k}$ while $\omega\simeq\mathrm{k}^{2}$ for small
$\mathrm{k}$. And this behavior is very close to the fermionic dispersion
curves obtained from the method of hard thermal loop (HTL) in thermal
field theory, that is \cite{key-46},

\begin{align}
\omega\left(\mathrm{k}\right) & \simeq m_{f}\pm\frac{1}{3}\mathrm{k}+\frac{1}{3m_{f}}\mathrm{k}^{2},\ \mathrm{k}\ll1;\nonumber \\
\omega\left(\mathrm{k}\right) & \simeq\mathrm{k},\ \mathrm{k}\gg1,
\end{align}
where $m_{f}$ is the effective mass generated by the medium effect
of fermion as, 

\begin{equation}
m_{f}=\sqrt{\frac{C_{F}}{8}}g_{\mathrm{YM}}T.
\end{equation}
Here $g_{\mathrm{YM}}$ refers to the Yang-Mills coupling constant
and $C_{F}$ is a constant suggested to be $C_{F}=4/3$ for fundamental
quarks or $C_{F}=1$ for electron. While the hard thermal loop approximation
\cite{key-46} displays two branches of the dispersion curves for
$\omega>0$ and $\omega<0$, it is out of reach for the present holographic
works with spinor e.g. \cite{key-41,key-42,key-44,key-45}.

\begin{figure}
\begin{centering}
\includegraphics[scale=0.27]{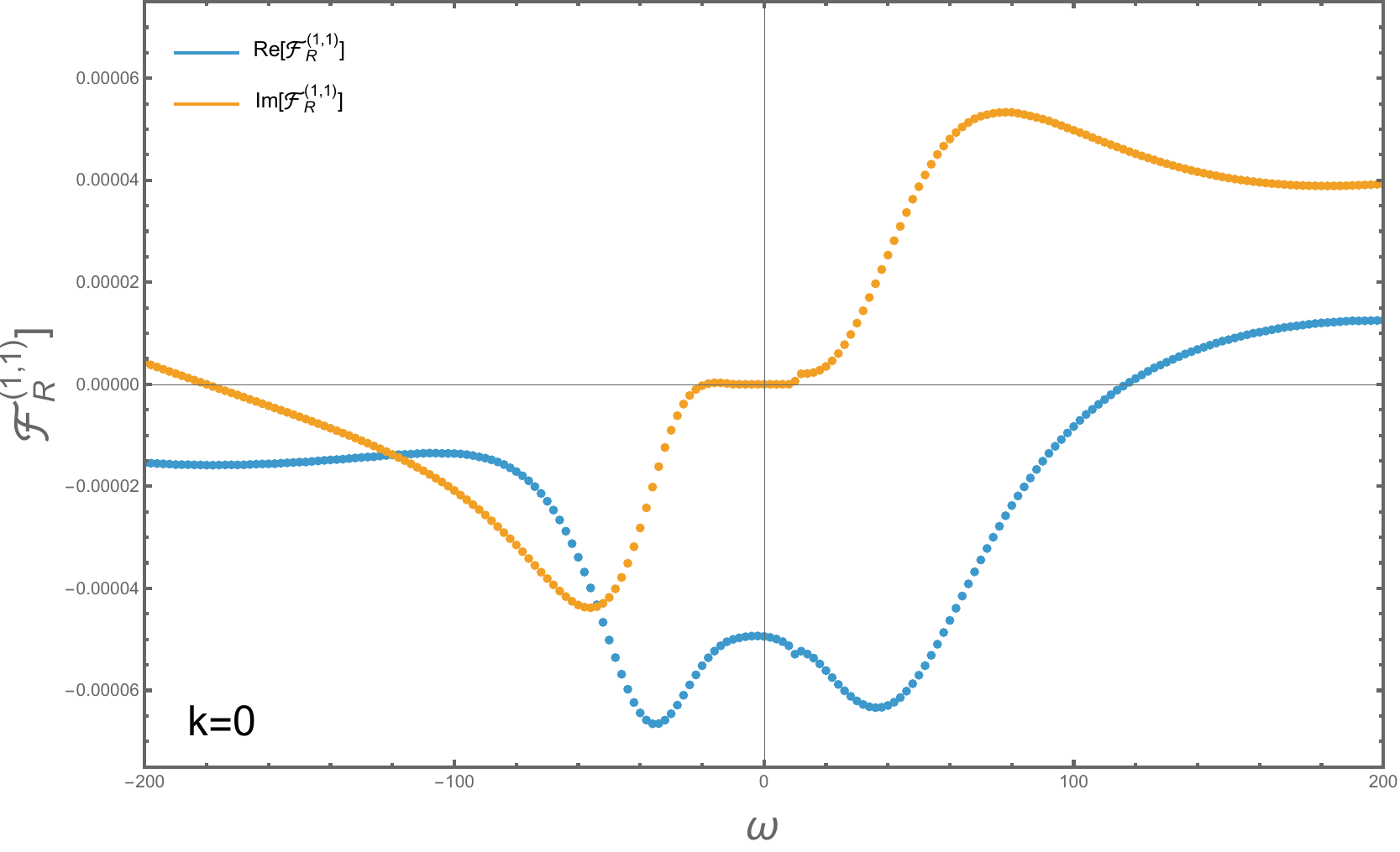}\includegraphics[scale=0.27]{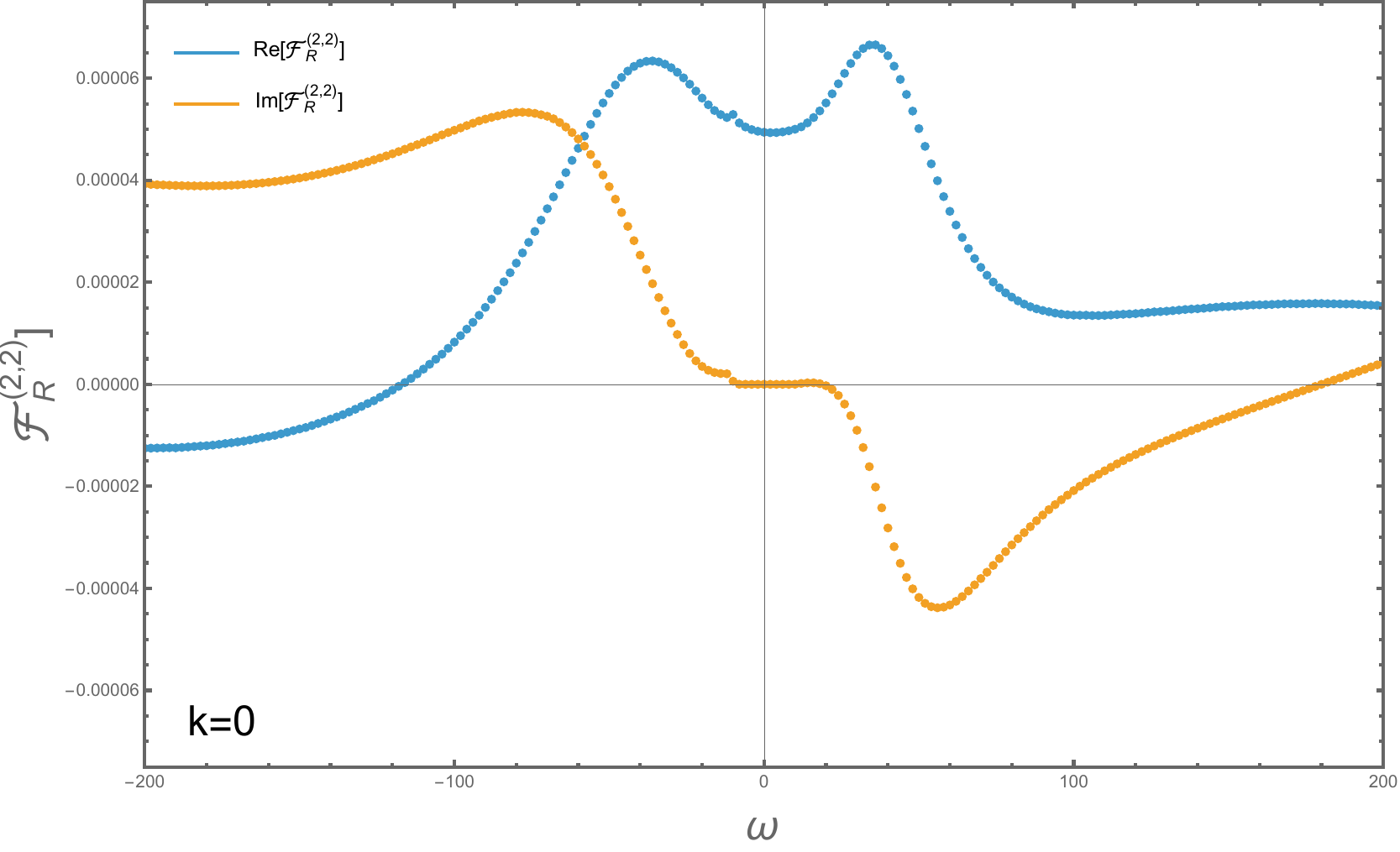}
\par\end{centering}
\begin{centering}
\includegraphics[scale=0.27]{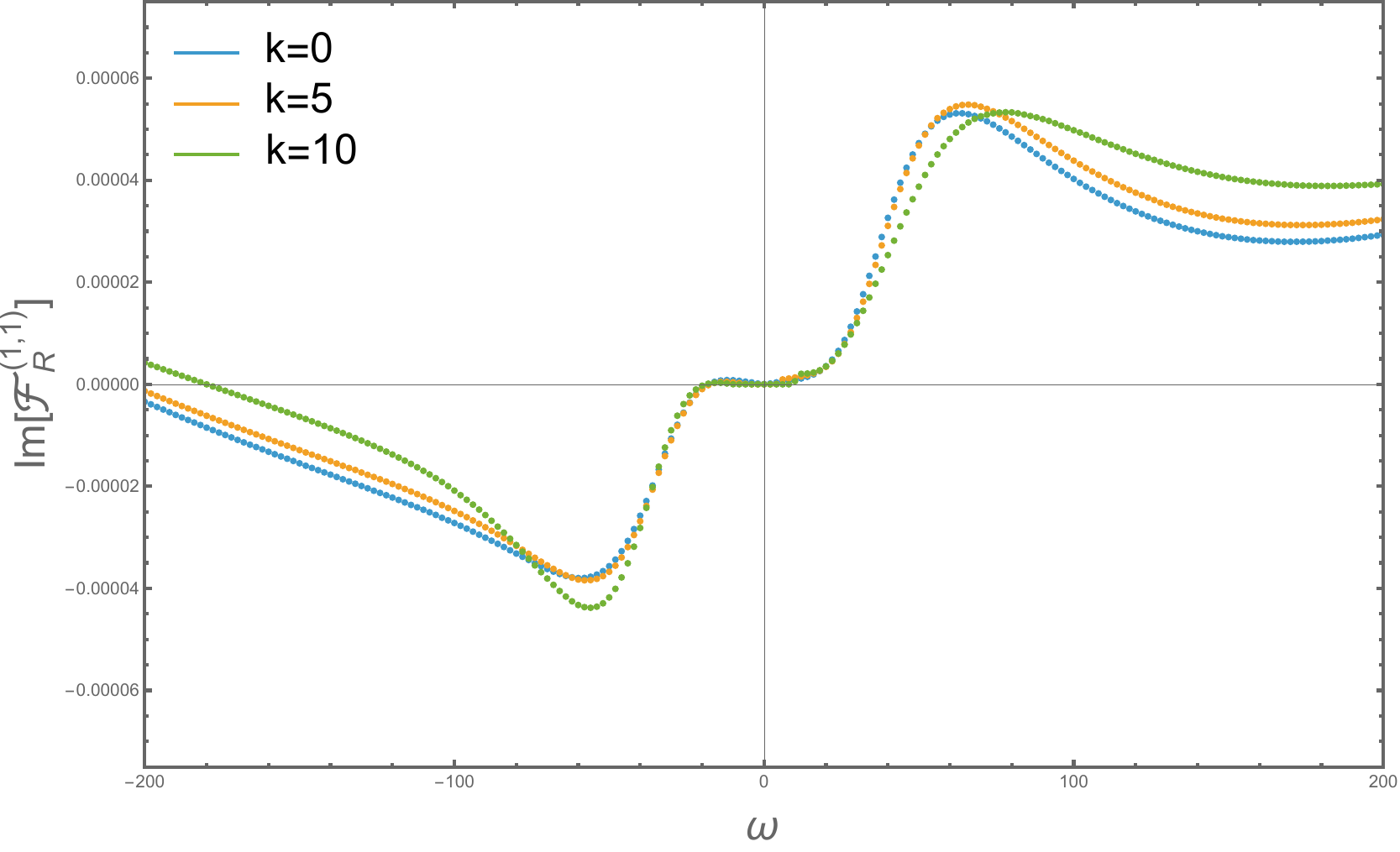}\includegraphics[scale=0.26]{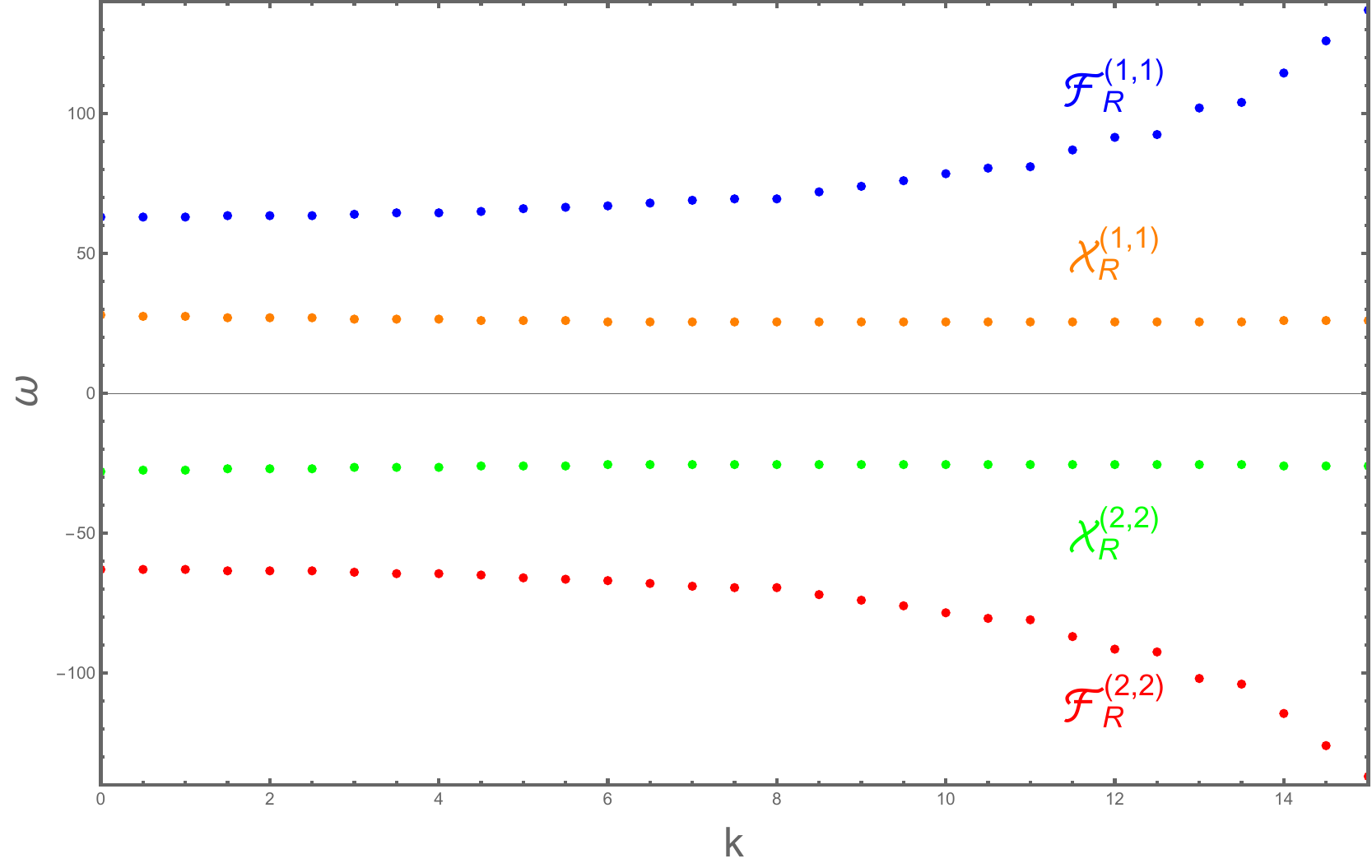}
\par\end{centering}
\caption{\label{fig:7}The first order Green function and the associated dispersion
curves in the deconfined phase.}

\end{figure}

Finally, let us investigate the instanton dependence in the spectral
function. The above numerical calculation illustrates that the spectral
function basically holds when the instanton charge increases, since
the dependence of the instanton density $Q$ in (\ref{eq:3.44}) is
not dominant at both boundary and horizon. Nevertheless, to see explicitly
the dependence of $Q$ in the spectral function, let us consider the
situation that the instanton density is sufficiently small i.e. $Q\ll1$.
And we assume the Green function can be decomposed as a series of
$Q$,

\begin{align}
G_{R} & =\mathcal{G}_{R}+Q\mathcal{F}_{R}+\mathcal{O}\left(Q^{2}\right),\nonumber \\
\xi_{\left(h\right)} & =X_{\left(h\right)}+QY_{\left(h\right)}+\mathcal{O}\left(Q^{2}\right).\label{eq:3.47}
\end{align}
where $\mathcal{G}_{R}$ is zero-th order Green function given by
solving (\ref{eq:3.44}) with $Q=0$ and $\mathcal{F}_{R}$ is first
order Green function which is given through (\ref{eq:3.35}) as,

\begin{equation}
\mathcal{F}_{R}^{\left(h,h\right)}=\left(-1\right)^{h}\lim_{\zeta\rightarrow\infty}\zeta Y_{\left(s\right)}.
\end{equation}
The equation of motion for $Y_{\left(h\right)}$ can be obtained by
plugging (\ref{eq:3.47}) into (\ref{eq:3.44}) which is

\begin{equation}
-\left(\frac{1}{\rho}+\frac{2\Lambda_{l}}{\zeta}\right)Y_{\left(h\right)}+X_{\left(h\right)}\left[\frac{2\ln\left(1-\frac{\rho^{2}}{r^{2}}\right)}{\rho^{3}\left(\rho^{2}-2r^{2}\right)}+\frac{2R^{2}}{z\rho}\left(\mathrm{k}+\frac{\omega}{1-\frac{\rho^{2}}{r^{2}}}Y_{\left(h\right)}\right)\right]-Y_{\left(h\right)}^{\prime}=0.\label{eq:3.49}
\end{equation}
And this equation can be solved numerically with the boundary condition
for $Y_{\left(h\right)}$ as $Y_{\left(h\right)}|_{r=r_{H}}=0$ which
is reduced from the infalling boundary condition $\xi_{\left(h\right)}|_{r=r_{H}}=\left(-1\right)^{h}i$.
Afterwards, the numerical evaluation for the first order Green function
$\mathcal{F}_{R}$ is plotted in Figure \ref{fig:7}. 
It shows that the onset mass increases by the momentum which leads
to a parallel dispersion curve additional to those from the first
order Green function. While there are two branches to each dispersion
curve with this approximation and it is close to the calculation from
the HTL, this approximation is very slight due to our numerical calculation.
Obviously, for $Q=0$, the approximated dispersion curve disappears.

\section{The interaction of the lowest baryon and meson}

In this section, let us investigate the interaction of the lowest
fermion and gauge boson as the lowest baryon and meson respectively
on the worldvolume of the D7-branes. The spectrum of the meson on
the flavor can be reviewed in the appendix which is also useful in
this section. As our concern is the interaction of hadrons, the following
analysis is valid in the confined phase only since quarks in deconfined
geometry can not form a hadron \cite{key-28,key-a5} in this model.
Keeping this in mind, let us focus on the trilinear Yukawa-like couplings
of the fermion and boson given in the action (\ref{eq:3.1}), since
both $\mathrm{L}_{\mathrm{D7}}$ and $\Gamma_{\mathrm{D7}}$ include
a piece linear in the gauge field strength $f$ as,

\begin{align}
\Gamma_{\mathrm{D7}} & =\gamma^{4}\left[1-\frac{1}{2}\bar{\gamma}\Gamma^{\alpha\beta}\left(2\pi\alpha^{\prime}\right)f_{\alpha\beta}+\mathcal{O}\left(f^{2}\right)\right],\nonumber \\
\mathrm{L}_{\mathrm{D7}} & =-\gamma^{4}\bar{\gamma}\Gamma^{\alpha}\left(2\pi\alpha^{\prime}\right)f_{\alpha\beta}g^{\beta\delta}D_{\delta}+\mathcal{O}\left(f^{2}\right),\label{eq:4.1}
\end{align}
where $\alpha,\beta,\delta$ runs over the D7-brane. Inserting (\ref{eq:4.1})
into action (\ref{eq:3.1}), we can obtain the trilinear coupling
terms as the lowest interaction terms involving mesons in the action
as,

\begin{align}
S_{int} & =S_{int}^{\left(1\right)}+S_{int}^{\left(2\right)},\nonumber \\
S_{int}^{\left(1\right)} & =-i\frac{T_{7}}{2}\left(2\pi\alpha^{\prime}\right)\int d^{8}x\sqrt{-g}e^{-\phi}\bar{\Psi}\frac{1}{2}\gamma^{4}\Gamma^{\alpha\beta}f_{\alpha\beta}\Gamma^{\gamma}D_{\gamma}\Psi,\nonumber \\
S_{int}^{\left(2\right)} & =-\frac{T_{7}}{2}\left(2\pi\alpha^{\prime}\right)\int d^{8}x\sqrt{-g}e^{-\phi}\bar{\Psi}\left(1-\gamma^{4}\right)\gamma^{4}\bar{\gamma}\Gamma^{\alpha}f_{\alpha\beta}g^{\beta\gamma}D_{\gamma}\Psi.\label{eq:4.2}
\end{align}
Follow the dimensional reduction in Section 3.2, then pick up the
confined metric (\ref{eq:2.10}), the actions presented in (\ref{eq:4.2})
can be written as, 

\begin{align}
S_{int}^{\left(1\right)} & =i\int d^{3}xd\zeta d\Omega_{4}\bar{\psi}\gamma^{\mu}\partial_{\mu}\pi\left(W_{1}\gamma^{\nu}\partial_{\nu}+W_{2}\gamma^{4}\partial_{4}+W_{3}\gamma^{4}+W_{4}\right)\psi,\nonumber \\
S_{int}^{\left(2\right)} & =-\frac{1}{2}\int d^{3}xd\zeta\bar{\psi}\left(\gamma^{4}-1\right)\left[P_{1}\gamma^{\mu}\partial_{\mu}\pi\partial_{4}-P_{1}\gamma^{4}\partial_{\mu}\pi+P_{2}\gamma^{\mu}\partial_{\mu}\pi\right]\psi.\label{eq:4.3}
\end{align}
where the potential functions dependent on $x^{4}\equiv\zeta$ are
collected as,

\begin{align}
W_{1} & =\frac{T_{7}}{2}\left(2\pi\alpha^{\prime}\right)d^{\left(0\right)}\sqrt{-\frac{g}{g_{44}g_{xx}g_{S^{5}}}}e^{-5\phi/4}\frac{R}{r},\nonumber \\
W_{2} & =\frac{T_{7}}{2}\left(2\pi\alpha^{\prime}\right)d^{\left(0\right)}\sqrt{-\frac{g}{g_{44}g_{xx}g_{S^{5}}}}e^{-5\phi/4}\frac{\rho}{R},\nonumber \\
W_{3} & =\frac{T_{7}}{2}\left(2\pi\alpha^{\prime}\right)d^{\left(0\right)}\sqrt{-\frac{g}{g_{44}g_{xx}g_{S^{5}}}}\left[\frac{e^{-5\phi/4}}{R}\left(-\frac{2\zeta}{\rho}+\frac{2\rho}{\zeta}+\frac{3\zeta}{2r}\frac{dr}{d\rho}+\frac{7}{8}\zeta\frac{d\phi}{d\rho}\right)-i\frac{3}{4}\frac{\rho}{R}e^{-\phi/4}\partial_{\zeta}C_{0}\right],\nonumber \\
W_{4} & =\frac{T_{7}}{2}\left(2\pi\alpha^{\prime}\right)d^{\left(0\right)}\sqrt{-\frac{g}{g_{44}g_{xx}g_{S^{5}}}}\left(\frac{1}{2}\frac{e^{-9\phi/4}}{R}+\frac{\rho}{R\zeta}e^{-5\phi/4}\Lambda_{l}^{\pm}\right),\nonumber \\
P_{1} & =T_{7}\left(2\pi\alpha^{\prime}\right)d^{\left(0\right)}\sqrt{-\frac{g}{g_{44}g_{xx}g_{S^{5}}}}e^{-\phi},\nonumber \\
P_{2} & =T_{7}\left(2\pi\alpha^{\prime}\right)d^{\left(0\right)}\sqrt{-\frac{g}{g_{44}g_{xx}g_{S^{5}}}}\frac{\zeta e^{-\phi}}{R^{2}}\left(\frac{1}{2}\frac{dr}{d\rho}+\frac{r}{8}\frac{d\phi}{d\rho}\right),\label{eq:4.4}
\end{align}
and we have imposed the decomposition of $f_{\alpha\beta}$ given
(\ref{eq:A15}) by picking up the lowest mesonic scalar identified
as $\varphi^{\left(0\right)}\equiv\pi$. Afterwards, let us further
recall the decomposition of $\psi$ given in (\ref{eq:3.13}) and
pick its lowest baryonic fermion as $\psi_{\pm}\equiv\psi_{\pm}^{\left(0\right)},f_{\pm}\equiv f_{\pm}^{\left(0\right)}$,
the actions in (\ref{eq:4.3}) can be rewritten as the effective coupling
terms as,

\begin{align}
S_{int}= & \int d^{3}x\partial_{\mu}\pi\bigg[\frac{ig_{1}}{M_{KK}}\left(\psi_{+}^{\dagger}\bar{\sigma}^{\mu}\sigma^{\nu}\partial_{\nu}\psi_{-}-\psi_{-}^{\dagger}\sigma^{\mu}\bar{\sigma}^{\nu}\partial_{\nu}\psi_{+}\right)\nonumber \\
 & +g_{2}\psi_{+}^{\dagger}\bar{\sigma}^{\mu}\psi_{+}-g_{3}\psi_{-}^{\dagger}\sigma^{\mu}\psi_{-}+g_{4}\left(\psi_{+}^{\dagger}i\partial^{\mu}\psi_{-}-\psi_{-}^{\dagger}i\partial^{\mu}\psi_{+}\right)\bigg],\label{eq:4.5}
\end{align}
where the coupling constants $g_{1,2,3,4}$ are functions of the instanton
density $Q$ which can be evaluated by the potentials given in (\ref{eq:4.4})
as, 
\begin{table}
\begin{centering}
\begin{tabular}{|c|c|c|c|c|}
\hline 
$Q=0$ & $g_{1}$ & $g_{2}$ & $g_{3}$ & $g_{4}$\tabularnewline
\hline 
\hline 
$\lambda^{-\frac{1}{4}}N_{c}^{-\frac{1}{2}}M_{KK}^{-1}$ & $1.65$ & $12.08$ & $-8.92$ & $1.46$\tabularnewline
\hline 
\end{tabular}
\par\end{centering}
\caption{\label{tab:3}The coupling constants of trilinear Yukawa-like terms
at $Q=0$ in the unit of $\lambda^{-\frac{1}{4}}N_{c}^{-\frac{1}{2}}M_{KK}^{-1}$.
The $\lambda=g_{s}N_{c}$ is the dimensionless 't Hooft coupling constant.}
 
\end{table}

\begin{align}
g_{1}\left(Q\right) & =\int d\zeta f_{+}W_{1}f_{-},\nonumber \\
g_{2}\left(Q\right) & =\int d\zeta f_{+}\left(W_{2}\partial_{4}+W_{3}+W_{4}\right)f_{+}-\int d\zeta\left(f_{+}P_{1}\partial_{4}f_{+}+f_{+}P_{2}f_{+}\right),\nonumber \\
g_{3}\left(Q\right) & =\int d\zeta f_{-}\left(W_{2}\partial_{4}+W_{3}-W_{4}\right)f_{-}-\int d\zeta\left(f_{+}P_{1}\partial_{4}f_{+}+f_{+}P_{2}f_{+}\right),\nonumber \\
g_{4}\left(Q\right) & =\int d\zeta f_{+}P_{1}f_{-}.\label{eq:4.6}
\end{align}

\begin{figure}
\begin{centering}
\includegraphics[scale=0.35]{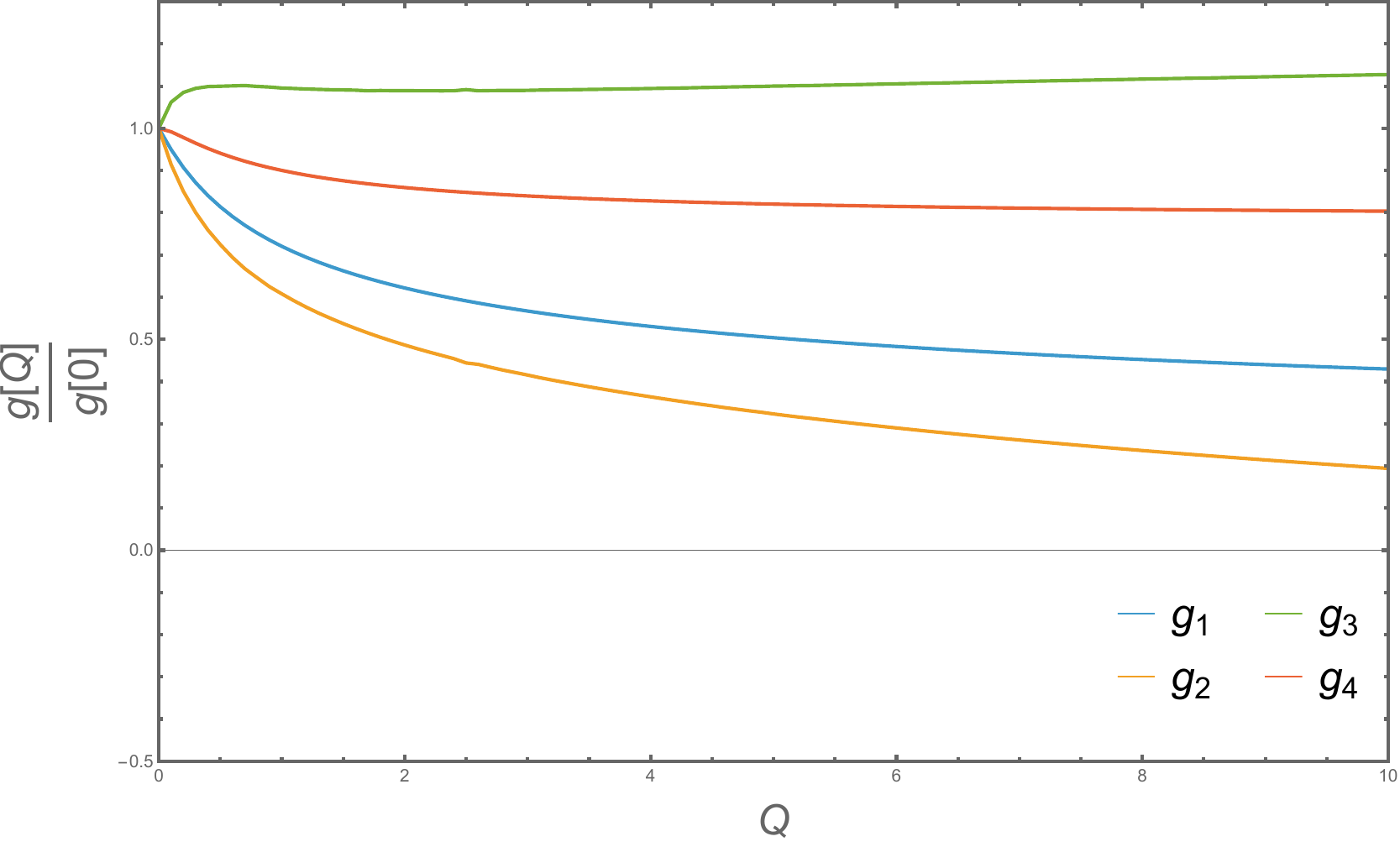}
\par\end{centering}
\caption{\label{fig:8}The ratios of $g\left(Q\right)/g\left(0\right)$ as
functions of $Q$. $g\left(Q\right)$ refers to the coupling constants
$g_{1,2,3,4}$.}
\end{figure}

Therefore, all the coupling constants can be evaluated numerically
where the associated results are given in Table \ref{tab:3} and in
particular, their dependences on the instanton density $Q$ are illustrated
numerically in Figure \ref{fig:8}. We can see that all the coupling
constants converge to a finite value at large $Q$, which implies
the amplitude of baryonic fermion and mesonic scalar basically holds
when the instanton density $Q$ increase. In addition, as the fermionic
fields are expected to be baryonic, there are $N_{c}$ quarks in $N_{c}$
possible insertions of the fermion bilinear. In large $N_{c}$, as
we have mentioned, we can rescale the fermionic field by $\psi_{+,-}^{\left(n\right)}\rightarrow\sqrt{N_{c}}\psi_{+,-}^{\left(n\right)}$
to include the contributions of the $N_{c}$ quarks in a baryonic
field. In this sense, all the coupling constants presented in (\ref{eq:4.6})
are proportional to $N_{c}^{1/2}$, which agrees with that the coupling
constant of baryon-meson interaction takes order of $N_{c}^{1/2}$
in large $N_{c}$ QCD \cite{key-40}. We finally note that action
(\ref{eq:4.5}) could be P violated if we identify ``$\pm$'' to
be the parity of the fermion.

\section{Summary and discussion}

In this work, we investigate the flavored worldvolume theory in the
top-down approach of D3/D7 model with homogenous D(-1)-branes. In
the gauge-gravity duality, D(-1)-branes are holographic instantons
which can violate the CP symmetry in the dual theory. Taking into
account the double Wick-rotation and dimensional reduction, we obtain
the confined and deconfined geometry in this model, which refers respectively
to the bubble and black brane solution. In the confined geometry,
we construct a 3d holographic theory of QCD involving baryon, meson
with instantons or CP violation. And the spectrum of the fermion is
obtained by decomposing the spinor on the worldvolume of the D7-brane
which is consistent quantitatively with the confined correlation function.
Moreover, our numerical evaluation illustrates that the mass spectrum
decreases slightly by the instanton density $Q$ and almost holds
at large $Q$. So the CP violation may not affect evidently the mass
of the hadrons predicted by this model. Interestingly, when the baryon
vertex as a D5-brane with $N_{c}$ open strings is introduced, the
worldvolume fermion as a gauge-invariant operator of $U\left(N_{c}\right)$
further takes the baryon number, therefore it can be identify to a
baryon in the confined geometry. In this sense, we compare the mass
of the lowest baryon and meson in our model with the associated experimental
data, then find they are basically close. Afterwards, we demonstrate
how to obtain the effective interaction terms of baryonic fermion
and mesonic boson, and evaluate the coupling constants by using the
data from the spectroscopy on the D7-brane. In the large $N_{c}$
limit, we find all the coupling constants are proportional to $N_{c}^{1/2}$
which agrees with the evaluation from the large N field theory \cite{key-40}
and it may not be necessary to preserve the CP symmetry in interaction
terms. So employing it to analyze the hadron decay with CP violation
additional to the CKM mechanism may be possible in the future works.
Besides, in the deconfined geometry, although the wrapped configuration
of a baryon vertex (D5-brane) may not exist, we also work out the
fermionic correlation function by using the holographic prescription
and interpret the fermion as plasmino \cite{key-44,key-45}. The dispersion
curve obtained from the deconfined correlation function basically
covers the results from HTL approximation \cite{key-46}. In summary,
this work is a systematical study about baryon, meson involving instantons
or CP violation through gauge-gravity in the top-down approach.

Let us give some more comments here to finalize this work. First,
the approach of the D3/D7 model with instantons is in fact based on
the IIB string theory, thus only two parameters $\lambda,M_{KK}$
living in our theory, which corresponds respectively to the string
coupling constant and the confining scale, are free to fit the experimental
data. Therefore ``less parameters'' is a characteristic feature
of the top-down approach different from the bottom-up approaches and
phenomenological models. Second, in the deconfined geometry, as we
have seen that most holographic approach can not work out the two
branches in the fermionic dispersion curve e.g. \cite{key-44,key-45},
the reason might be that the background geometry we used corresponds
to the ideal fluid where the dissipative terms are absent. In this
sense, re-evaluate the holographic fermionic correlation function
with dissipative terms in the background geometry (e.g. \cite{key-47})
may be interesting in the future works. Third, although we obtain
the second branch in the fermionic dispersion curve (displayed in
Figure \ref{fig:7}) by solving the first order equations for the
correlation function in (\ref{eq:3.49}), it is due to the presence
of the instantons. And this separation in the dispersion curve may
be caused by the instanton-induced interaction with spin as it is
discussed in \cite{key-6,key-7}, so the gauge-gravity duality may
provide a powerful way to investigate it since confirming the correlation
function of baryon with instantons is quite difficult in QFT approach.
Last but not least, in the confined geometry, while the dual field
theory in our model is definitely QCD{\scriptsize 3} in the large $N_{c}$
limit, we believe that most of the evaluation in this work is less
dependent on the dimension. The reason is that the spectroscopy and
the coupling constants of both fermion and boson are determined by
the basis functions given in (\ref{eq:3.13}) and (\ref{eq:A6}) whose
eigen equations are basically independent on the dimension given (\ref{eq:3.17})
and (\ref{eq:A10}). Therefore, as a holographic approach, our current
evaluation may also be suitable for 4d QCD.

\section*{Acknowledgements}

This work is supported by the National Natural Science Foundation
of China (NSFC) under Grant No. 12005033, the Fundamental Research
Funds for the Central Universities under Grant No. 3132025192.

\section*{Appendix: The mesonic spectrum from the DBI action of the flavor
brane}

\begin{figure}
\begin{centering}
\includegraphics[scale=0.27]{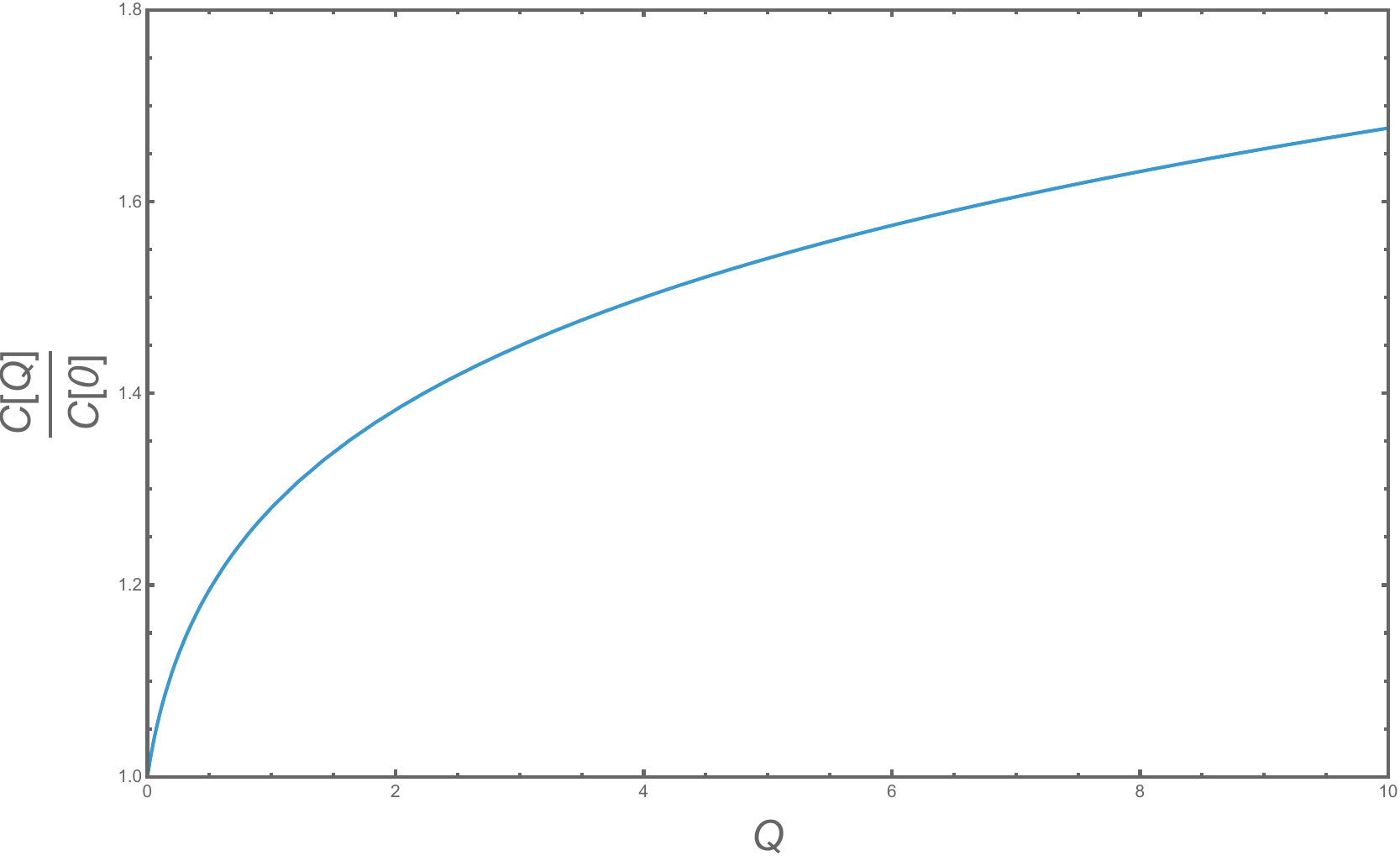}\includegraphics[scale=0.27]{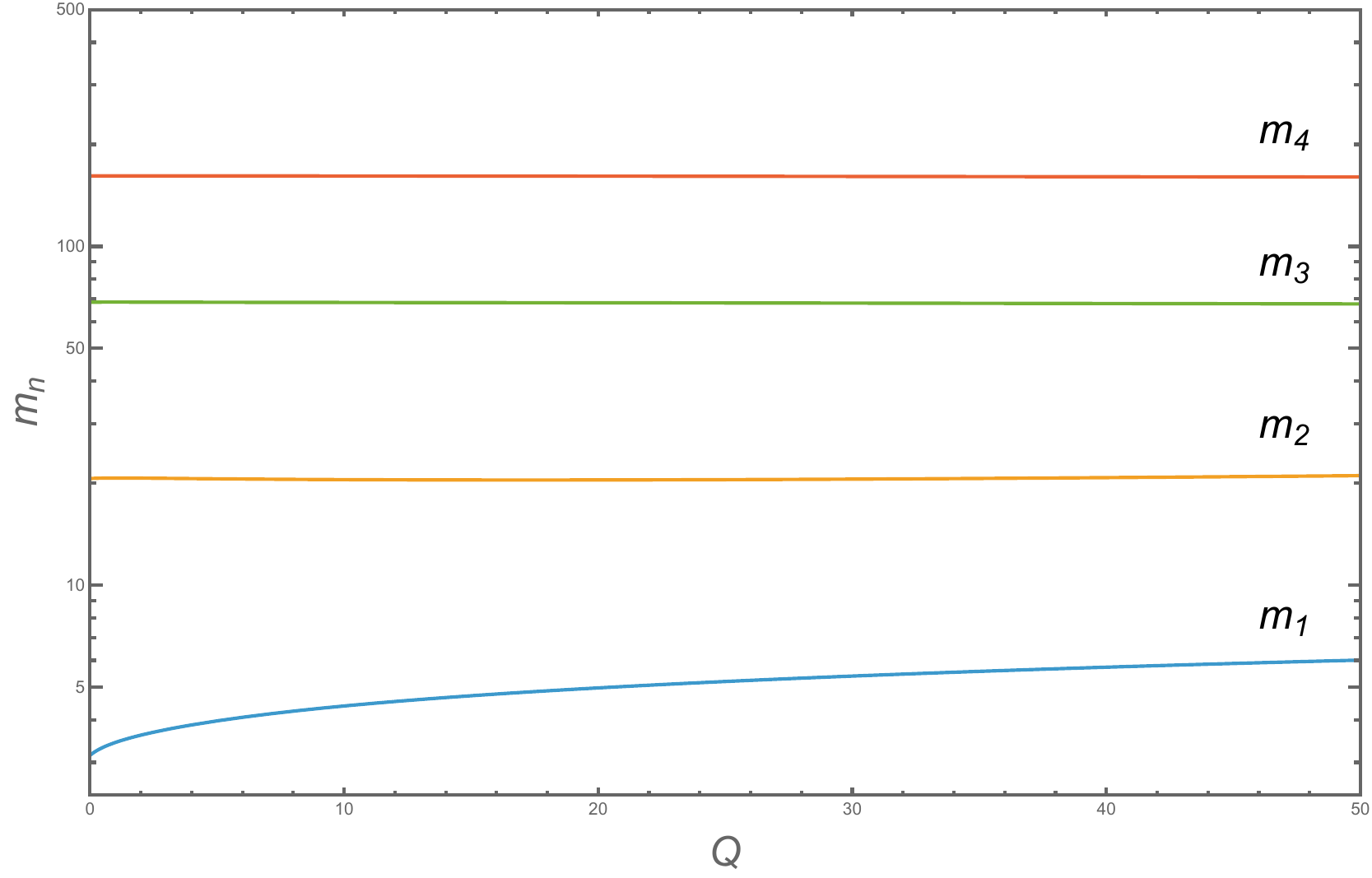}
\par\end{centering}
\caption{\textbf{\label{fig:A1}Left: }The ratio of $C\left(Q\right)/C\left(0\right)$
as a function of $Q$. \textbf{Right:} The mesonic spectrum as a function
of $Q$.}
\end{figure}

In this appendix, let us outline the spectrum of the mesonic gauge
boson on the worldvolume of the (flavor) D7-brane by using the Dirac-Born-Infeld
(DBI) action. Since mesons are confined states, the following analysis
is valid only in the confined geometry given in (\ref{eq:2.5}). Recall
the DBI action for a D7-brane given as,

\begin{align}
S_{\mathrm{D7}} & =-T_{\mathrm{D7}}\int d^{8}xe^{-\phi}\sqrt{-\det\left(g_{\alpha\beta}+2\pi\alpha^{\prime}f_{\alpha\beta}\right)},\nonumber \\
 & \simeq-T_{\mathrm{D7}}\int_{\mathrm{D7}}d^{3}xd\zeta d\Omega_{4}e^{-\phi}\sqrt{-g}\left[1+\frac{1}{4}\left(2\pi\alpha^{\prime}\right)^{2}g^{\alpha\gamma}g^{\beta\delta}f_{\alpha\beta}f_{\gamma\delta}+\mathcal{O}\left(f^{4}\right)\right],\tag{A1}\label{eq:A1}
\end{align}
where $f_{\alpha\beta},g_{\alpha\beta}$ refers respectively to the
gauge field strength and induced metric on the flavor brane. To obtain
the standard kinetic form for vector and scalar meson, the quadratic
term in the action is necessary, so we need,

\begin{equation}
S_{\mathrm{D7}}=-\frac{1}{4}\left(2\pi\alpha^{\prime}\right)^{2}T_{\mathrm{D7}}\int d^{3}xd\zeta d\Omega_{4}e^{-\phi}\sqrt{-g}g^{\alpha\gamma}g^{\beta\delta}f_{\alpha\beta}f_{\gamma\delta},\tag{A2}\label{eq:A2}
\end{equation}
where the indices $\alpha,\beta,\delta$ run over the D7-brane. Plugging
the induced metric on the D7-brane given in (\ref{eq:2.11}) into
(\ref{eq:A2}) and assuming the non-vanished components of the gauge
field are ($x^{4}\equiv\zeta$),
\begin{equation}
A_{\alpha}\left(x,\zeta\right)=\left\{ A_{\mu}\left(x,\zeta\right),A_{4}\left(x,\zeta\right)\right\} ,\tag{A3}
\end{equation}
the action (\ref{eq:A2}) can be written as,

\begin{equation}
S_{\mathrm{D7}}=-\frac{1}{4}\int d^{3}xd\zeta U_{1}\left(\eta^{\mu\rho}\eta^{\nu\sigma}f_{\mu\nu}f_{\rho\sigma}+2U_{2}f_{\mu4}f_{\nu4}\right),\tag{A4}\label{eq:A4}
\end{equation}
where $f_{\mu\nu}=\partial_{\mu}A_{\nu}-\partial_{\nu}A_{\mu}$ is
the kinetic term of the gauge field and

\begin{equation}
U_{1}\left(\xi\right)=\frac{2}{3\pi^{2}}\frac{N_{c}}{M_{KK}}e^{2\phi}\frac{\zeta^{4}r_{KK}}{\rho^{5}r},U_{2}\left(\xi\right)=\frac{r^{2}}{4r_{KK}^{2}}M_{KK}^{2}\rho^{2}.
\end{equation}
Then we can decompose the gauge field by a series of complete functions
as,

\begin{equation}
A_{\mu}\left(x,\zeta\right)=\sum_{n=1}^{\infty}B_{\mu}^{\left(n\right)}\left(x\right)b^{\left(n\right)}\left(\zeta\right),A_{4}\left(x,\zeta\right)=\sum_{n=1}^{\infty}\varphi^{\left(n\right)}\left(x\right)d^{\left(n\right)}\left(\zeta\right).\tag{A6}\label{eq:A6}
\end{equation}
Inserting (\ref{eq:A6}) into (\ref{eq:A4}), it leads to,

\begin{align}
S_{\mathrm{D7}}= & -\frac{1}{4}\sum_{n,m=1}^{\infty}\int d^{3}xd\zeta U_{1}\bigg\{\left[\partial_{\mu}B_{\nu}^{\left(n\right)}-\partial_{\nu}B_{\mu}^{\left(n\right)}\right]\left[\partial^{\mu}B^{\nu\left(m\right)}-\partial^{\nu}B^{\mu\left(m\right)}\right]b^{\left(n\right)}b^{\left(m\right)}\nonumber \\
 & +2U_{2}\left[\partial_{\mu}\varphi^{\left(n\right)}d^{\left(n\right)}\left(\zeta\right)-B_{\mu}^{\left(n\right)}\partial_{\zeta}b^{\left(n\right)}\right]\left[\partial^{\mu}\varphi^{\left(m\right)}d^{\left(m\right)}-B^{\mu\left(n\right)}\partial_{\zeta}b^{\left(n\right)}\right]\bigg\}.\tag{A7}\label{eq:A7}
\end{align}
In order to obtain the standard form of the kinetic term, the following
normalization conditions are required for the basis functions as,

\begin{align}
\int d\zeta U_{1}b^{\left(n\right)}b^{\left(m\right)} & =\delta^{mn},\nonumber \\
\int d\zeta U_{1}U_{2}\partial_{\zeta}b^{\left(n\right)}\partial_{\zeta}b^{\left(m\right)} & =m_{n}\delta^{mn},\tag{A8}\label{eq:A8}
\end{align}
with $m_{n}=\lambda_{n}M_{KK}$ and

\begin{equation}
\int d\zeta U_{1}U_{2}d^{\left(n\right)}d^{\left(m\right)}=\delta^{mn}.\tag{A9}
\end{equation}
So these conditions reduce to the solution for the $d^{\left(n\right)}$
, $\partial_{\zeta}b^{\left(n\right)}$ and the eigen equations for
the basis function $b^{\left(n\right)}$ ($n\geq1$) as,

\begin{align}
\partial_{\zeta}\left[U_{1}U_{2}\partial_{\zeta}b^{\left(n\right)}\right] & =\lambda_{n}M_{KK}U_{1}b^{\left(n\right)},\nonumber \\
d^{\left(n\right)} & =m_{n}^{-1}\partial_{\zeta}b^{\left(n\right)}.\tag{A10}\label{eq:A10}
\end{align}
Notice that since the basis functions $b^{\left(n\right)}$'s have
to satisfy the Dirichlet boundary condition, there is an additional
function $d^{\left(0\right)}$ orthogonal to all the $d^{\left(n\right)}$'s
for $n\geq1$ if we chose 
\begin{equation}
d^{\left(0\right)}=\frac{C}{U_{1}U_{2}}.\tag{A11}
\end{equation}
So, one can verify that

\begin{equation}
\int d\zeta U_{1}U_{2}d^{\left(0\right)}d^{\left(n\right)}=Cm_{n}^{-1}\int d\zeta\partial_{\zeta}b^{\left(n\right)}=0,\ n\geq1.\tag{A12}
\end{equation}
Further using the normalization condition,

\[
r_{KK}^{-1}\int d\zeta U_{1}U_{2}d^{\left(0\right)}d^{\left(0\right)}=1,\tag{A13}
\]
we obtain

\[
C=\left(r_{KK}^{-1}\int d\zeta U_{1}U_{2}\right)^{-1/2},\tag{A14}
\]
where the constant $C$ must depend on the density of instanton $Q$.
And dependence of $Q$ is numerically evaluated in Figure \ref{fig:A1}.
Keeping all the above in mind, it means we need to shift the start
point in the summary given in $A_{4}$ in (\ref{eq:A6}) as,

\begin{equation}
A_{4}\left(x,\zeta\right)=\varphi^{\left(0\right)}\left(x\right)d^{\left(0\right)}\left(\zeta\right)+\sum_{n=1}^{\infty}\varphi^{\left(n\right)}\left(x\right)d^{\left(n\right)}\left(\zeta\right).\tag{A15}\label{eq:A15}
\end{equation}

Afterwards, perform the gauge transformation as,

\begin{equation}
V_{\mu}^{\left(n\right)}\left(x\right)=B_{\mu}^{\left(n\right)}-m_{n}^{-1}\partial_{\mu}\varphi^{\left(n\right)},\tag{A16}
\end{equation}
then substitute $\partial_{\zeta}b^{\left(n\right)}$ for $d^{\left(n\right)}$
and insert (\ref{eq:A8}) (\ref{eq:A15}) into (\ref{eq:A7}), the
action in (\ref{eq:A7}) becomes the standard form of vector and scalar
field as,

\begin{equation}
S_{\mathrm{D7}}=-\sum_{n=1}^{\infty}\int d^{3}x\left[\frac{1}{4}W_{\mu\nu}^{\left(n\right)}W^{\mu\nu\left(n\right)}+\frac{1}{2}m_{n}V_{\mu}^{\left(n\right)}V^{\mu\left(n\right)}\right]-\frac{1}{2}\int d^{3}x\partial_{\mu}\varphi^{\left(0\right)}\partial^{\mu}\varphi^{\left(0\right)},\tag{A17}
\end{equation}
where

\begin{equation}
W_{\mu\nu}=\partial_{\mu}V_{\nu}-\partial_{\nu}V_{\mu}.\tag{A18}
\end{equation}
The mass spectrum for the vector meson $V_{\mu}$ can be obtained
by solving the eigen equations for $b^{\left(n\right)}$ which is
given in (\ref{eq:A10}), and the numerical result is given in Table
\ref{tab:A1} and Figure \ref{fig:A1}. 
\begin{table}
\begin{centering}
\begin{tabular}{|c|c|c|c|c|}
\hline 
$m_{n}$ & $n=1$ & $n=2$ & $n=3$ & $n=4$\tabularnewline
\hline 
\hline 
$Q=0$ & 3.15 & 20.64 & 68.43 & 161.5\tabularnewline
\hline 
$Q=1$ & 3.43 & 20.71 & 68.47 & 161.51\tabularnewline
\hline 
\end{tabular}
\par\end{centering}
\caption{\label{tab:A1}The spectrum of the mesonic boson in the unit of $M_{KK}=1$
with various $Q$.}

\end{table}
Our numerical results indicate that the mass spectrum is less dependent
on the instanton density $Q$. Beside, since the confined phase described
by (\ref{eq:2.11}) is valid below the scale $M_{KK}$, we note that
only the massless scalar $\varphi^{\left(0\right)}$ arises strictly
in the low-energy theory. Finally, {]} the action of the D7-brane
also contains CS terms as,

\begin{equation}
S_{\mathrm{CS}}=\frac{\mu_{7}}{8!}\int C_{8}+\frac{\mu_{7}}{2!}\left(2\pi\alpha^{\prime}\right)^{2}\int f\wedge f\wedge C_{4}+\mathrm{fermionic\ terms},\tag{A19}
\end{equation}
which are of order $\mathcal{O}\left(f^{2}\right)$ and are CP violated
terms.

\end{document}